\documentstyle[12pt]{article}
\newcommand{\be}{\begin{equation}}
\newcommand{\ee}{\end{equation}}
\newcommand{\bea}{\begin{eqnarray}}
\newcommand{\eea}{\end{eqnarray}}
\newcommand{\nn}{\nonumber \\}
\newcommand{\p}[1]{(\ref{#1})}
\newcommand{\lb}{\label}

\begin{document}
\begin{titlepage}
\begin{flushright}
hep-th/0611247\\
November 2006
\end{flushright}
\vskip 0.6truecm

\begin{center}
{\Large\bf Gauging ${\cal N}{=}4$ supersymmetric mechanics II:
\vspace{0.2cm}

${\bf (1,4,3)}$ models from the ${\bf (4,4,0)}$ ones}
\vspace{1.5cm}

{\large\bf F. Delduc$\,{}^a$, E. Ivanov$\,{}^b$,}\\
\vspace{1cm}

{\it a)ENS Lyon, Laboratoire de Physique, 46, all\'ee d'Italie,}\\
{\it 69364 LYON Cedex 07, France}\\
{\tt francois.delduc@ens-lyon.fr}
\vspace{0.3cm}

{\it b)Bogoliubov  Laboratory of Theoretical Physics, JINR,}\\
{\it 141980 Dubna, Moscow region, Russia} \\
{\tt eivanov@theor.jinr.ru}\\

\end{center}
\vspace{0.2cm}
\vskip 0.6truecm  \nopagebreak

\begin{abstract}
\noindent Exploiting the gauging procedure developed by us in {\tt hep-th/0605211},
we study the relationships between the models of ${\cal N}{=}4$ mechanics
based on the off-shell multiplets
${\bf (4,4,0)}$ and ${\bf (1,4,3)}\,$. We make use of the off-shell ${\cal N}{=}4$, $d=1$
harmonic superspace approach as most adequate for treating this circle of problems.
We show that the most general sigma-model type
superfield action of the multiplet ${\bf (1,4,3)}$ can be obtained in a few non-equivalent ways
from the ${\bf (4,4,0)}$ actions invariant under certain three-parameter
symmetries, through gauging the latter by the appropriate non-propagating gauge
multiplets. We discuss in detail the gauging of both the Pauli-G\"ursey $SU(2)$ symmetry
and the abelian three-generator shift symmetry. We reveal the ${\bf (4,4,0)}$ origin of
the known mechanisms of generating potential terms for the multiplet ${\bf (1,4,3)}\,$,
as well as of its superconformal properties. A new description of this multiplet in terms
of unconstrained harmonic analytic gauge superfield is proposed. It suggests, in particular,
a novel mechanism of generating the ${\bf (1,4,3)}$ potential terms via coupling
to the fermionic off-shell ${\cal N}{=}4$ multiplet ${\bf (0,4,4)}\,$.
\end{abstract}
\vspace{0.7cm}

\newpage

\end{titlepage}

\section{Introduction}

An extended $d=1$ supersymmetry possesses some notable features which
are not shared by its higher-dimensional counterparts. In view of the distinguished
role of supersymmetric quantum mechanics, both as the appropriate simplified
``laboratory'' for studying various  aspects of supersymmetric quantum field
theories and as a theory providing superextensions of some intrinsically one-dimensional
systems, it is of importance to fully understand these specific features
of the $d=1$ supersymmetry and their dynamical manifestations in the corresponding
models of supersymmetric mechanics. One of these peculiarities
is the so-called ``1D automorphic duality''
\cite{GR0}-\cite{GR} which relates off-shell $d=1$ supermultiplets with the same number
of physical fermions and different divisions of the set of bosonic fields into
physical and auxiliary components (see also \cite{PT,T,GenSm}).
The procedure generating the multiplets with
a greater number of auxiliary fields from those with the lesser number
and the procedure inverse to it can be referred to as the ``reduction'' and
``oxidation'', respectively \cite{I}\footnote{In an obvious analogy
with the nomenclature used in the context of the space-time dimensional reduction.}.
The non-linear versions of this duality were considered in \cite{IL}-\cite{root}.
Using these dualities, the relations between different supersymmetric mechanics models
can be studied. In particular, it was argued in \cite{root} that various models
of ${\cal N}{=}4$ supersymmetric mechanics based on the off-shell multiplets
with 4 physical fermions can be obtained, through the reduction procedure, from
the models based on the ``root'' \cite{Root} off-shell multiplet with the field content
${\bf (4, 4,0)}$ \cite{GR0,Pol,IL,IKL1,Root}(hereafter, the abbreviating
${\bf (n_1, n_2 = n_1 +n_3, n_3)}$ stands for
the off-shell multiplet with ${\bf n_1}$ physical bosons, ${\bf n_2}$
physical fermions and ${\bf n_3}$ auxiliary bosonic fields).

In most of the above studies, the reduction procedure was accomplished at the
component level and ``by hands'': basically by treating the time derivative of
some initial physical bosonic field as a new auxiliary field. Recently, we
proposed a superfield version of this procedure ensuring the manifest off-shell supersymmetry
at all steps \cite{I}. The process of reduction was shown to
amount to gauging some isometries of the superfield actions of the multiplet with
the maximal number of the physical bosonic fields (${\bf (1, 1,0)}$, ${\bf (2, 2, 0)}$ and
${\bf (4,4,0)}$ in the ${\cal N}{=}1$,  ${\cal N}{=}2$  and ${\cal N}{=}4$ cases) by
a ``topological'' gauge multiplet. The characteristic property
of the latter (specific just for the $d=1$ case) is that its only surviving
component in the Wess-Zumino gauge is the bosonic
``gauge field''. The residual gauge freedom is always realized in such a way that
it can be used to kill one or few original bosonic fields. After fully fixing
the gauge freedom in this way, one is left with the new off-shell multiplet in which the place
of ``killed'' bosonic fields is occupied by the former gauge fields which possess
no kinetic terms and so are auxiliary. Thus the supersymmetric mechanics of
$d=1$ supermultiplets with one or another numbers of auxiliary fields naturally arises
upon fixing a gauge in a coupled system of the ``extreme'' multiplet  having
no auxiliary fields at all and a ``topological'' supermultiplet which
gauges one or another isometry realized on the extreme multiplet. Besides choosing the
Wess-Zumino gauge (which basically corresponds to the component consideration of refs.
\cite{GR0}-\cite{GenSm}, \cite{root}) one is free to choose another,
manifestly supersymmetric gauge in which
the whole reduction procedure can be performed in terms of superfields. This possibility
to accomplish the reduction in a manifestly supersymmetric superfield fashion is one of
the merits of the gauging approach. It can be regarded as an efficient tool
of deducing off-shell superfield descriptions of $d=1$ supersymmetric
mechanics systems, starting from the system associated with the basic (``root'')
multiplet. In most cases, the potential terms of the resulting multiplet
are generated by Fayet-Iliopoulos terms of the gauge superfield and/or
the gauge-covariantized Wess-Zumino-like terms of the ``root'' multiplet.
The inverse ``oxidation'' procedure amounts to constraining the gauge multiplet
to be ``pure gauge'' and so eliminating it altogether.

In \cite{I} we concentrated on the ${\cal N}{=}4$ mechanics and the superfield
reduction from the ${\cal N}{=}4$ root multiplet ${\bf (4, 4, 0)}$ to the multiplet
${\bf (3, 4, 1)}$ (both its linear \cite{ISmi,BP,IKL0} and non-linear \cite{IL,IKL1}
versions).
As a by-product we constructed a new non-linear superfield version of the multiplet
${\bf (4,4,0)}$ in ${\cal N}{=}4$, $d=1$ harmonic superspace
(see also \cite{Pol}, \cite{UnP}-\cite{burg}). Some
further examples of our gauging procedure
leading to the ${\cal N}{=}4$ multiplets ${\bf (0, 4, 4)}$ and ${\bf (1,4,3)}$
were also considered. The last case corresponds to gauging non-abelian
$SU(2)$ symmetry realized on the multiplet ${\bf (4,4,0)}$.

In \cite{I} we limited our study to a particular case of such gauging,
with the free action of the multiplet ${\bf (4,4,0)}$ as the input.
In the present paper we consider the most general
situation. We start from the most general $SU(2)$ invariant ${\bf (4,4,0)}$
action in the harmonic ${\cal N}{=}4, d{=}1$ superspace and show
that its $SU(2)$ gauging generates the generic sigma-model
type superfield action of the multiplet ${\bf (1,4,3)}$ \cite{IKLe,IKPa}. Also, we show that
the latter can be equally reproduced by gauging some other three-parameter
isometries admitting a realization on the multiplet ${\bf (4,4,0)}\,$.
As such one can choose three commuting shift isometries.
We discuss the ${\bf (4,4,0)}$ origin of various mechanisms
of generating potential terms of the multiplet ${\bf (1, 4, 3)}$
and show how the superconformally invariant actions of the latter
can be reproduced within the gauging procedure from the superconformal
${\bf (4,4,0)}$ actions. As a by-product we find out a new description
of the multiplet ${\bf (1,4,3)}$ in terms of unconstrained
harmonic analytic gauge prepotential. This description suggests
a new mechanism of generating potential terms of the multiplet ${\bf (1,4,3)}$
via coupling it to the off-shell fermionic ${\cal N}{=}4$ multiplet
${\bf (0,4,4)}\,$. Also, using this formulation, off-shell couplings of
the multiplet ${\bf (1,4,3)}$ to the ${\bf (3,4,1)}$ one
and some extra ${\bf (4,4,0)}$ multiplets can be easily constructed.
Finally, we discuss the description of the ``mirror''
${\bf (1,4,3)}$ multiplet (with a different $SU(2)$ assignment
of the component fields) in the ${\cal N}{=}4, d{=}1$ harmonic superspace
and present a simple coupling of the mirror multiplet to the initial
${\bf (1,4,3)}$ multiplet.
\setcounter{equation}{0}

\setcounter{equation}{0}
\section{${\cal N}{=}4\,, \; d{=}1$ harmonic superspace}

\subsection{Basics}
Because the ${\cal N}{=}4$, $d{=}1$ harmonic superspace (HSS) plays the central role
in our construction, we start by recollecting the basics of this approach following
refs. \cite{HSS,HSS1} and \cite{IL,I}.

The ${\cal N}{=}4, d{=}1$ superspace is defined as the following coordinate set
\be
z = (t,\theta_{i}\,\; \bar\theta^{i})\,,\quad \bar\theta^{i} = \overline{(\theta_{i})}\,.\lb{N4}
\ee
The covariant spinor derivatives are defined as
\begin{eqnarray}
&& D^{i}=\frac{\partial}{\partial \theta_{i}}+i\bar\theta^{i}\partial_{t}\,,\quad
\bar D_{i}=\frac{\partial}{\partial \bar\theta^{i}}+i\theta_{i}\partial_{t} = -\overline{(D^i)}\,, \nn
&& \{D^{i},\bar D_{j}\}=2i\delta^i_j\partial_{t}\,, \;\{D^{i}, D^{j}\}
= \{\bar D_{i},\bar D_{j}\} =0\,.
\end{eqnarray}
The ${\cal N}{=}4,\; d{=}1$ harmonic superspace (HSS) in the {\it central} basis
is the following extension of \p{N4}
\be
(z, u) = (t,\;\theta_{i}\,,\; \bar\theta^{i}\,,\; u^\pm_i)\,.
\ee
Here $u^\pm_i \in SU(2)_A/U(1)\,$ are the $SU(2)_A$ harmonic variables:
\be
u^{-}_i = \overline{(u^{+i})}\,, \quad u^{+i}u^-_{i}=1\; \Leftrightarrow \;
u^+_iu^-_k - u^+_ku^-_i = \varepsilon_{ik}\,. \lb{CompL}
\ee
The coordinates of ${\cal N}{=}4, \;d{=}1$ HSS in the {\it analytic} basis are
\begin{equation}
\left( t_{A}=t-i(\theta^+\bar\theta^-+\theta^-\bar\theta^+)\,,\;
\theta^\pm=\theta^{i}u^\pm_{i}\,,\; \bar\theta^\pm=\bar\theta^{i}u^\pm_{i}\,, \;u^\pm_k \right).
\end{equation}
The analytic subspace of HSS is defined as
\begin{equation}
(t_{A},\theta^+,\bar\theta^+, u^\pm_{i})\equiv (\zeta, u).
\end{equation}
It is closed under the ${\cal N}{=}4$ supersymmetry
\be
\delta t_A = -2i\left(\epsilon^iu^-_i \bar\theta^+ - \bar\epsilon^iu^-_i\theta^+\ \right),
\quad \delta \theta^+ =\epsilon^iu^+_i\,, \;\delta \bar\theta^+ =
\bar\epsilon^iu^+_i\,, \;\delta u^\pm_i = 0\,,
\label{SUSYan}
\ee
and is real with respect to the generalized conjugation $\widetilde{\;\;}$ \cite{HSS}
\be
\widetilde{t_A} = t_A\,, \; \widetilde{\theta^\pm} = \bar\theta^\pm\,, \; \widetilde{\bar\theta^\pm} = -\theta^\pm\,, \; \widetilde{u^\pm_i} = u^{\pm i}\,, \;
\widetilde{u^{\pm i}} = - u^\pm_i\,. \lb{Tilde}
\ee

In the central basis $(z\,,\, u^{\pm i})$, the harmonic derivatives
and the harmonic projections of spinor derivatives are defined by
\be
D^{\pm\pm} = \partial^{\pm\pm} = u^{\pm}_i\frac{\partial}{\partial u^{\mp}_i}\,,
\quad D^\pm=u^\pm_{i}D^{i},\quad \bar D^\pm=u^\pm_{i}\bar D^{i}\,. \lb{Cderiv}
\ee
In the analytic basis, the same spinor and harmonic derivatives read
\begin{eqnarray}
&& D^+=\frac{\partial}{\partial\theta^{-}}\,,\,\, \bar D^+=
-\frac{\partial}{\partial\bar\theta^-}\,,\,\,
D^{-}=-\frac{\partial}{\partial \theta^{+}}+2i\bar\theta^{-}\partial_{t_{A}}\,,\,\,
\bar D^{-}=\frac{\partial}{\partial \bar\theta^{+}}+2i\theta^{-}\partial_{t_{A}}\,,\nn
&& D^{++}=\partial^{++}-2i\theta^+\bar\theta^+\partial_{t_{A}}
+\theta^+\frac{\partial}{\partial\theta^-}
+ \bar\theta^+\frac{\partial}{\partial\bar\theta^-}\,, \nn
&& D^{--}=\partial^{--}-2i\theta^-\bar\theta^-\partial_{t_{A}}
+\theta^-\frac{\partial}{\partial\theta^+}
+  \bar\theta^-\frac{\partial}{\partial\bar\theta^+}\,.
\end{eqnarray}
They satisfy the following non-zero (anti)commutation relations
\begin{eqnarray}
&&[D^{\pm\pm},D^\mp]=D^\pm\,,\,\, [D^{\pm\pm},\bar D^\mp]=\bar D^\pm\,,\,\,
\{D^+,\bar D^-\}=-\{D^-,\bar D^+\}=2i\partial_{t_{A}}\,,  \nn
&& [D^{++},D^{--}]= D^{0}\,, \quad [D^0, D^{\pm\pm}] =
\pm 2 D^{\pm\pm}\,, \lb{DharmAl}\\
&& D^0 = u^+_{i}\frac{\partial}{\partial u^+_{i}}-u^-_{i}
\frac{\partial}{\partial u^-_{i}}+
\theta^+\frac{\partial}{\partial \theta^+}
+\bar\theta^+\frac{\partial}{\partial \bar\theta^+}
-\theta^-\frac{\partial}{\partial \theta^-}
-\bar\theta^-\frac{\partial}{\partial \bar\theta^-}\,.\label{Dalg}
\end{eqnarray}
The derivatives $D^+\,$, $\bar D^+$ are short in the analytic basis,
whence it follows that one can define analytic ${\cal N}{=}4$ superfields
$\Phi^{(q)}(\zeta, u)$
\be
D^+\Phi^{(q)} = \bar D^+\Phi^{(q)}=0 \quad \Rightarrow \quad \Phi^{(q)}
= \Phi^{(q)}(\zeta, u)\,,\lb{AnalPhi}
\ee
where $q$ is the external harmonic $U(1)$ charge. This Grassmann harmonic
analyticity is preserved by the harmonic derivative $D^{++}\,$:
when applied to $\Phi^{(q)}(\zeta, u)$, this derivative yields an analytic
${\cal N}{=}4, d{=}1$ superfield of charge $(q+2)$.

The measures of integration over the full HSS and its analytic subspace
are given, respectively, by
\begin{eqnarray}
&& \mu_H = dudtd^4\theta=dudt_{A}(D^-\bar D^-)(D^+\bar D^+)=\mu_{A}^{(-2)}(D^+\bar D^+),\nn
&& \mu_{A}^{(-2)}=dud\zeta^{(-2)}
=dudt_{A}d\theta^+d\bar\theta^+=dudt_{A}(D^-\bar D^-)\,. \label{measures}
\end{eqnarray}
The integration over the harmonic two-sphere $S^2 = \{u^+_i, u^-_k\}$ is normalized so that
\be
\int du\, 1 = 1\,,
\ee
and the integral of any other irreducible monomial of the harmonics
is vanishing \cite{HSS,HSS1}.

\subsection{Multiplet $q^{+a}$ and its symmetries}
In this paper we shall deal with the multiplet ${\bf (4,4,0)}$ which is described
by a doublet analytic superfield $q^{+a}(\zeta,u)$ of charge $1$
satisfying the harmonic constraint \footnote{For brevity, in what follows we frequently
omit the index ``A'' of $t_A$.}
\begin{equation}
D^{++}q^{+a}=0 \quad \Rightarrow \quad q^{+a}(\zeta,u)=f^{ia}(t)u^+_{i}+\theta^+\chi^{a}(t)
+\bar\theta^+\bar\chi^{a}(t)+2i\theta^+\bar\theta^+\partial_t f^{ia}(t)u^-_{i}.\label{coq}
\end{equation}
It satisfies the pseudoreality condition (see \p{Tilde})
\be
\widetilde{q^{+ a}} = - q^{+ a}\, \Rightarrow \, \overline{(f^{ia})} = \epsilon_{ab}\epsilon_{ik} f^{kb}\,, \; \overline{(\chi^a)} = \bar\chi_a\,. \lb{RealCond} \ee

The Grassmann analyticity conditions together with the harmonic constraint \p{coq}
imply that in the central basis
\bea
q^{+ a} = q^{ia}(t,\theta, \bar\theta)u^{+}_i\,, \;\;
D^{(i}q^{k)a} = \bar D^{(i}q^{k)a} = 0\,. \label{qconsC}
\eea

We may write a general off-shell action for the ${\bf (4,4,0)}$ multiplet as
\begin{equation}
S_{q}=\int dudtd^4\theta \,{\cal L}(q^{+a}, q^{-b}, u^\pm) , \quad
q^{-a}\equiv D^{--}q^{+a}. \lb{qact}
\end{equation}
After solving the constraint \p{coq} in the central basis of HSS,
the superfield $q^{\pm a}$
may be written in this basis as
$q^{\pm a}=u^{\pm}_{i}q^{ia}(t,\theta,\bar\theta)\,$, $D^{(i}q^{k) a}
= \bar D^{(i}q^{k)a} = 0$.
We then use the notation
\begin{equation}
L(q^{ia})=\int du\,{\cal L}(q^{+a}, q^{-b}, u^\pm),\quad S_{q}
=\int dtd^4\theta\, L(q^{ia})\,. \label{qact1}
\end{equation}
The free action is given by
\begin{equation}
S_{q}^{\mbox{\scriptsize free}}
= -\frac{1}{4}\,\int dudtd^4\theta\,(q^{+a}q^-_{a})
=\frac{i}{2}\int dud\zeta^{(-2)}\,(q^{+a}\partial_t q^+_{a}). \lb{Freeq}
\end{equation}

The action \p{qact} produces a sigma-model type action in components,
with two time derivatives on
the bosonic fields and one derivative on the fermions.
One can also construct an invariant which
in components yields a Wess-Zumino type action,
with one time derivative on the bosonic fields
(plus Yukawa-type fermionic terms). It is given by
the following general integral over
the analytic subspace
\be
S^{WZ}_q = \int du d\zeta^{(-2)}\, {\cal L}^{+2}(q^{+a}, u^\pm)\,. \lb{WZq}
\ee

The free action \p{Freeq} and constraint \p{coq} exhibit a number of symmetries.
Some of them can be extended to the interaction case, leading to certain
restrictions on the form of the general action \p{qact}. In terms of the component fields,
these symmetries become isometries of the target bosonic metric. We shall list here
all symmetries of this sort, having in mind to gauge some of them in the next Sections.
We will be interested only in those symmetries which commute with ${\cal N}{=}4$
supersymmetry and so can be gauged without passing to the local supersymmetry \cite{I}.
\vspace{0.3cm}

\noindent{\it Rotational isometries.} The free action \p{Freeq} and constraint \p{coq} are manifestly
invariant under the so-called Pauli-G\"ursey $SU(2)$ group
\be
\delta_{su(2)}\, q^{+ a} = \lambda^a_{\;\;b}\, q^{+ b}\,,
\quad \lambda^a_{\;\;a} = 0\,, \quad \overline{(\lambda^a_{\;\;b})}= -\lambda_a^{\;\;b}\,. \label{PG}
\ee
An arbitrary one-parameter subgroup of this group is singled out as
\be
\delta_{1}\, q^{+ a} = \lambda_1 \,c^a_{\;\;b}\, q^{+ b} \equiv \lambda_1\, {\cal I}_1(c)\, q^{+a} \,, \;\;{\cal I}_1(c) = c^{a}_{\;\;b}q^{+ b}\frac{\partial}{\partial q^{+ a}}\,,  \label{U1}
\ee
where $c^{(ab)}$ is an isotriplet of constants
\be
c^a_{\;\;a} = 0\,, \quad \overline{(c^{ab})}= c_{ab}\,.
\ee
\vspace{0.2cm}
One more isometry of this type is the target space dilatations
\be
\delta_2\, q^{+ a} = \lambda_2\, q^{+ a} \equiv \lambda_2\, {\cal I}_2\, q^{+ a}\,, \;\; {\cal I}_2 = q^{+ a}\frac{\partial}{\partial q^{+ a}}\,.
 \label{scale}
\ee
The constraint \p{coq} is obviously covariant under these rescalings, while the action
\p{Freeq} is not.
The simplest invariant action is of the sigma model type \cite{I}. The transformations \p{scale}
and \p{PG} commute with each other.
\vspace{0.2cm}

\noindent{\it Shift isometries.} The action \p{Freeq} (up to boundary terms) and constraint
\p{coq} are also invariant under the abelian shifts
\be
\mbox{(a)}\;\delta_3\, q^{+ a} =
\lambda_3\, u^{+a} \equiv \lambda_3\, {\cal I}_3\, q^{+ a}\,,\;\; {\cal I}_3 = u^{+ a}\frac{\partial}{\partial q^{+ a}}\,; \qquad \mbox{(b)} \;
\delta_4 q^{+a} =  \tilde{\lambda}^{(a}_{\;\;\;b)}\,u^{+ b}\,. \label{Shift1}
\ee
An arbitrary one-parameter subgroup of (\ref{Shift1}b) is singled out as
\be
\delta_5\, q^{+a} = \lambda_5\, b^{(a}_{\;\;\;d)}\, u^{+ d} \equiv
\lambda_5\,{\cal I}_5(b)\, q^{+ a}\,, \;\; {\cal I}_5(b) = b^{(a}_{\;\;\;d)}\, u^{+ d}\frac{\partial}{\partial q^{+ a}}\,,\label{Is5}
\ee
\vspace{0.2cm}
where $b^{(ab)}$ is some real constant isotriplet (in general, it is different
from $c^{(ab)}$). In what follows, without loss of generality, we normalize
all these isotriplets as in \cite{I}
\be
c^2 = c^{ab}c_{ab} = 2\,, \qquad b^2 = b^{ab}b_{ab} = 2\,.
\ee

The rotational and shift isometries, being realized on the same $q^{+ a}$,
do not commute with each other. Their closure is an extension of $SU(2)_{PG}$ by some
solvable subgroups. Below we list some subgroups with two and three
generators from this closure.
\vspace{0.2cm}

\noindent{\it Commuting subsets}
\bea
&& G_I = \{{\cal I}_1(c), {\cal I}_2 \}\,, \quad G_{II}
= \{{\cal I}_3, {\cal I}_5(b) \}\,, \quad G_{III} =
\{{\cal I}_5(c), {\cal I}_5(b) \}\,,\lb{Comm} \\
&& \qquad \; \;\left[ {\cal I}_1(c), {\cal I}_2\right] = \left[ {\cal I}_3, {\cal I}_5(b) \right] =
\left[ {\cal I}_5(c), {\cal I}_5(b) \right] = 0\,.
\eea

\noindent{\it Non-commuting subsets}
\bea
&& G_{IV} = \{{\cal I}_2, {\cal I}_3 \}\,, \quad G_{V} = \{{\cal I}_2, {\cal I}_5(b) \}\,,
\qquad \;\, \left[{\cal I}_3, {\cal I}_2 \right] = {\cal I}_3\,,
 \;\;
\left[{\cal I}_2, {\cal I}_5(b) \right] = -{\cal I}_5(b)\,, \label{25} \\
&& G_{VI} = \{{\cal I}_1(c), {\cal I}_3, {\cal I}_5 (c)\}\,, \qquad \;
\left[{\cal I}_1(c), {\cal I}_3 \right] = -{\cal I}_5(c)\,, \quad
\left[{\cal I}_1(c), {\cal I}_5 (c) \right]
= {\cal I}_3\,, \label{135} \\
&& G_{VII} = \{{\cal I}_1(c), {\cal I}_5(b), {\cal I}_5 (d)\}\,, \quad (c\cdot b) = (c\cdot d) = (b\cdot d)= 0\,, \quad
d^{ab} \equiv c^{(a}_{\;f}b^{fb)}\,,\nn
&& \qquad\quad \;\;\left[{\cal I}_1(c), {\cal I}_5(b) \right] = -{\cal I}_5(d)\,, \quad
\left[{\cal I}_1(c), {\cal I}_5 (d) \right]
= {\cal I}_5(b)\,, \label{136} \\
&& G_{VIII} = \{{\cal I}_2, {\cal I}_3, {\cal I}_5 (c)\}\,, \;\quad
\left[{\cal I}_2, {\cal I}_3 \right] = -{\cal I}_3\,, \quad
\left[{\cal I}_2, {\cal I}_5 (c) \right]
= -{\cal I}_5(c)\,, \label{137} \\
&& G_{IX} = \{{\cal I}_2, {\cal I}_5(c), {\cal I}_5 (b)\}\,,\;(c\cdot b) = 0\,, \;
\left[{\cal I}_2, {\cal I}_5(c) \right] = -{\cal I}_5(c)\,, \;
\left[{\cal I}_2, {\cal I}_5 (b) \right]
= -{\cal I}_5(b)\,. \label{138}
\eea

The subclasses of the general
$q$-actions \p{qact} and \p{WZq} revealing invariance under various subgroups listed above
will be defined in the next Sections, as far as necessary. In fact, some of the above subgroups coincide up to a redefinition of $q^{+ a}$ by some constant matrix (preserving the constraint \p{coq} and the free action \p{Freeq}). The full list of such isomorphisms is \be
G_{II} \sim G_{III}\,, \quad G_{IV} \sim G_{V}\,, \quad G_{VI} \sim G_{VII}\,, \quad G_{VIII} \sim G_{IX}\,. \label{Iso}
\ee

For further use, we also present the full structure of the closure of the rotational and shift isometries.
Denoting, in the proper basis, the generators of $SU(2)_{PG}$ \p{PG} by
$T_M$ $(M = 1,2,3)$, the
generator of target dilatations \p{scale} by $T$ and the generators of the singlet and triplet shifts
in \p{Shift1} by $F$ and $F_M$, we have
\bea
&&\left[ T_M\,, T_N\right] = i\varepsilon_{MNK}\,T_K\,, \quad
\left[ T_M, F_N\right] =
\frac{i}{2}\,\varepsilon_{MNK}\,F_K - i\delta_{MN} F\,, \quad
\left[ T_M\,, F \right] = \frac{i}{4} F_M\,, \nonumber \\
&& \left[ F_M\,, F_N\right] = \left[ F_M, F \right] = 0\,, \quad
\left[ T, T_M\right] = 0\,, \quad
\left[ T, F_M\right] = F_M\,, \quad \left[ T, F \right] = F\,. \label{Closure}
\eea
This algebra generates a subgroup of the Weyl group in 4-dimensional target Euclidean space,
with $F$ and $F_M$ forming the 4-translation operator and $SU(2)_{PG}$ being one of two $SU(2)$
factors of the rotation group $SO(4) \sim SU(2)\times SU(2)\,$\footnote{The second factor
is the automorphism $SU(2)$ acting on the doublet indices of harmonics $u^{\pm i}\,$, central basis
Grassmann variables and the left index of the superfield $q^{ia}$ in \p{qconsC}. Since
it does not commute with ${\cal N}{=}4$ supersymmetry, it cannot be gauged without turning on
the whole world-line ${\cal N}{=}4$ supergravity \cite{I}.}.

\subsection{``Topological'' gauge ${\cal N}{=}4$ superfield}
The ${\cal N}{=}4$, $d{=}1$ ``gauge multiplet'' is described by a charge
$2$ unconstrained analytic superfield
$V^{++}(\zeta,u)\,$ the gauge transformation of which in the abelian case reads
\begin{equation}
\delta V^{++}=D^{++}\Lambda\,,
\end{equation}
with $\Lambda(\zeta,u)$ being a charge zero unconstrained analytic
superfield parameter.
Using this gauge freedom, one can choose the Wess-Zumino gauge, in which
the gauge superfield becomes
\begin{equation}
V^{++}(\zeta,u)=2i(\theta^+\bar\theta^+)A(t),\quad\delta A(t)=
-\partial_{t}\Lambda_{0}(t),\,\,
\Lambda_0 = \Lambda(\zeta,u)\vert_{\theta = 0}\,. \lb{WZ}
\end{equation}
We observe that the ``gauge''
${\cal N}{=}4, d{=}1$ multiplet locally carries $(0 + 0)$ degrees of freedom
and so it is ``topological''.
Globally the field $A(t)$ can differ from a pure gauge, and this feature
allows for its treatment
as an auxiliary field in the ``unitary'' gauges.

As in the ${\cal N}{=}2, d{=}4$ HSS \cite{HSS,HSS1}, $V^{++}$ gauge-covariantizes
the analyticity-preserving
harmonic derivative $D^{++}\,$. Assume that the analytic superfield $\Phi^{(q)}$
is transformed under
some abelian gauge isometry as
\be
\delta_\Lambda \Phi^{(q)} = \Lambda\, {\cal I}\, \Phi^{(q)}\,,
\ee
where ${\cal I}$ is the corresponding generator. Then the harmonic derivative
$D^{++}$ is
covariantized as
\be
D^{++}\Phi^{(q)} \quad \Longrightarrow \quad {\cal D}^{++}\Phi^{(q)} =
(D^{++} - V^{++}\,{\cal I})\Phi^{(q)}\,.\lb{D++cov}
\ee
One can also define the second, non-analytic harmonic connection $V^{--}$
\be
{\cal D}^{--} = D^{--} - V^{--}\,{\cal I}\,, \quad \delta V^{--}
= D^{--} \Lambda\,. \lb{D--cov}
\ee
{}From the requirement of preserving the algebra of harmonic derivatives \p{DharmAl},
\be
[{\cal D}^{++}, {\cal D}^{--}] = D^0\,, \quad [D^0, {\cal D}^{\pm\pm}]
= \pm 2 {\cal D}^{\pm\pm}\,,
\ee
the well-known harmonic zero-curvature equation follows
\be
D^{++}V^{--} - D^{--}V^{++} = 0\,. \lb{Hzc}
\ee
It specifies $V^{--}$ in terms of $V^{++}$. One can also define
the covariant spinor derivatives
\be
{\cal D}^{-} = [{\cal D}^{--}, D^+] = D^- +(D^+V^{--})\, {\cal I}\,, \quad
\bar{\cal D}^{-} = [{\cal D}^{--}, \bar D^+] =
\bar D^- + (\bar D^+V^{--})\, {\cal I}\,, \lb{Spcov}
\ee
as well as the covariant time derivative ${\cal D}_t$:
\be
\{D^+, \bar{\cal D}^- \} = 2i {\cal D}_t\,, \quad {\cal D}_t
= \partial_t  - \frac{i}{2}(D^+\bar D^+ V^{--})\,{\cal I}\,. \lb{Vectcov}
\ee
The vector gauge connection
\be
V \equiv D^+\bar D^+ V^{--},\, \quad \delta V
= -2i\partial_{t_A} \Lambda\,, \lb{Vconn}
\ee
is an analytic superfield, $D^+ V = \bar D^+ V = 0\,$,
so ${\cal D}_t$ preserves the analyticity. In the WZ gauge \p{WZ} \be
V \quad \Longrightarrow \quad 2i\,A(t)\,. \lb{WZV}
\ee

We will exploit these relations and their non-abelian generalization
in next Sections.

\setcounter{equation}{0}
\section{Multiplet (1, 4, 3) from gauging $SU(2)_{PG}$}
In the full set of rotational and shift symmetries realized on $q^{+ a}$ there are
six different subgroups with three generators: $SU(2)_{PG}$ defined in \p{PG}, the
abelian 3-parameter translation set (\ref{Shift1}b) and four solvable subgroups $G_{VI}$
- $G_{IX}$ defined by eqs. \p{135} - \p{138} (modulo isomorphisms \p{Iso}). In this paper we shall limit our study to the first two options\footnote{See the concluding Sect. 5 for some comments on the remaining options.} and start with gauging $SU(2)_{PG}\,$. The simplest version of this gauging, with the free $q^{+a}$ action as the point of departure,
was already considered in \cite{I}.

\subsection{From (4, 4, 0) to (1, 4, 3)}
Let us gauge the $SU(2)_{PG}$ symmetry \p{PG} by substituting $\lambda^a_{\;b} \rightarrow
\Lambda^a_{\;b}(\zeta, u)\,$,
\be
\delta q^{+ a} = \Lambda^{a}_{\;\;\;b}q^{+ b}\,. \label{qtraN}
\ee
The constraint \p{coq} is covariantized to
\be
\nabla^{++}q^{+ a} \equiv D^{++}q^{+ a} - {V^{++}}^a_{\;b}q^{+ b}  = 0\,,\label{NAcoq}
\ee
where the traceless analytic gauge connection ${V^{++}}^a_{\;b}$ is transformed as
\be
\delta {V^{++}}^a_{\;b}=  D^{++}\Lambda^a_{\;b} + \Lambda^a_{\;c}{V^{++}}^c_{\;b}
- {V^{++}}^a_{\;c}\Lambda^c_{\;b}\,.\label{traN++}
\ee

The analytic superspace form of the free action \p{Freeq} is covariantized by replacing
\be
\partial_t q^{+ a} \quad \Rightarrow \quad \nabla_t q^{+a}
= \partial_t q^{+a} -\frac{i}{2} V^a_{\,\;b}q^{+ b}\,,
\ee
where
\bea
&& V^a_{\;\;b} = D^+\bar D^+ {V^{--}}^a_{\;b}\,, \label{Vdef} \\
&& D^{++} {V^{--}}^a_{\;b} -  D^{--}{V^{++}}^a_{\;b}
- {V^{++}}^a_{\;c}{V^{--}}^c_{\;b} + {V^{--}}^a_{\;c}{V^{++}}^c_{\;b} =0\,, \label{H0cond} \\
&& \delta {V^{--}}^a_{\;b}=  D^{--}\Lambda^a_{\;b} + \Lambda^a_{\;c}{V^{--}}^c_{\;b}
- {V^{--}}^a_{\;c}\Lambda^c_{\;b}\,. \label{traN--}
\eea
The equivalent form of the action \p{Freeq} in the central basis is covariantized by replacing
\be
q^{- a} = D^{--}q^{+ a} \quad \Rightarrow \quad \hat{q}{}^{- a} \equiv \nabla^{--}q^{+ a} =
D^{--}q^{+ a} - {V^{--}}^a_{\;b}q^{+ b}\,. \label{Repl}
\ee
It is straightforward to check the identity of both forms of the covariantized free action.

Using the constraint \p{NAcoq} and the harmonic zero curvature condition \p{H0cond},
it is easy to check that
\be
\left[ \nabla^{++}, \nabla^{--} \right] = D^0\,, \quad \nabla^{++} \hat{q}{}^{- a} = q^{+ a}\,,
\quad \nabla^{--} \hat{q}{}^{- a} = 0\,. \label{Rel}
\ee

The subclass of general $q^{+}$ actions \p{qact} enjoying gauge $SU(2)_{PG}$ symmetry
is defined as follows
\be
S_{gauge} = \int dt d^4\theta du {\cal L}(q^{+ a}\hat{q}^{-}_a, u^\pm)\,. \label{GaugeL}
\ee
Taking into account the relations \p{NAcoq}, \p{Rel}, the $SU(2)_{PG}$ invariant
\be
J \equiv q^{+ a}\hat{q}^{-}_a \label{defJ}
\ee
is the only gauge invariant quantity which one can construct. Also, it is
easy to show that
\be
D^{++} J = 0\,,
\ee
whence it follows that $J$ does not depend on harmonics in the central basis.
Therefore, without loss of generality, one can neglect the harmonic integral in \p{GaugeL}
together with the dependence on the explicit harmonics in ${\cal L}(J, u^\pm)\,$. Thus,
the most general gauge invariant action is obtained, via the replacement \p{Repl}, from the most general globally $SU(2)_{PG}$ invariant action
\be
S_{PG} = \int dt d^4\theta\,  {\cal L}\left(q^{+ a}{q}^{-}_a\right) =
\int dt d^4\theta\,  {\cal L}\left(\frac{1}{2}q^{ia}{q}_{ia}\right). \label{SPG}
\ee

As was already shown in \cite{I}, the $SU(2)_{PG}$ gauging of the multiplet $q^{+a}$
gives rise to one sort of the ${\cal N}{=}4, d{=}1$ supermultiplet ${\bf (1,4,3)}$,
in such a way that three physical bosonic components of $q^{+ a}$ become purely gauge
while $V^{++(ab)}$ supplies three auxiliary degrees of freedom. In \cite{I} this
was demonstrated in the WZ gauge
\bea
&&{V^{++}}^a_{\;b} = 2i \theta^+\bar\theta^+ A^{a}_{\,\;b}(t)\,, \;
{V^{--}}^a_{\;b} = 2i\theta^-\bar\theta^-\,A^{a}_{\;\;b}(t)\,,\;
V^a_{\;\;b} = D^+\bar D^+ {V^{--}}^a_{\;b} =
2iA^{a}_{\;\;b}(t)\,,\label{WZ2} \\
&&\delta_r A^{a}_{\,\;b} =
-\partial_t{\Lambda_{(0)}}^a_{\;b}+ {\Lambda_{(0)}}^a_{\;c}A^{c}_{\;b}
- A^{a}_{\;c}{\Lambda_{(0)}}^c_{\;b}\,, \quad \delta_r f^{ia}
=  {\Lambda_{(0)}}^a_{\;c} f^{ic}\,,  \label{WZ1}
\eea
where $f^{ia}(t)$ is the first component of $q^{+ a}$, $f^{ia}u^+_i = q^{+ a}|_{\theta = 0}\,$.
In this gauge, the solution of the covariantized constraint \p{NAcoq}
is obtained from the solution \p{coq}
just by the replacement
\be
\partial_t f^{ia} \;\Rightarrow \;\nabla_t f^{ia} = \partial_t f^{ia}
+ A^a_{\;\;b}f^{i b}\,.
\ee
Splitting $f^{ia}$ as
\be
f^{ia} = \varepsilon^{ia} \frac{1}{\sqrt{2}}\,f + f^{(ia)}\,, \lb{Split}
\ee
and assuming that $f$ has a non-vanishing constant vacuum part,
$f = <f> + \ldots\,, <f> \neq 0$, one
observes that the symmetric part in \p{Split} can be fully gauged away
by the residual $SU(2)$ gauge
freedom
\be
f^{ia} \quad \Rightarrow \quad \varepsilon^{ia} \frac{1}{\sqrt{2}}\,f\,. \label{ResG}
\ee
So one ends up with the fields $f(t), \psi^{ia}(t), A^{(ab)}\,$, which is just the off-shell
field content of the multiplet ${\bf (1,4, 3)}\,$. Note that in this gauge
the only manifest $SU(2)$ symmetry is the diagonal one in the product $SU(2)_A\times SU(2)_{PG}\,$.
It plays the role of automorphism $SU(2)$ group. As usual in WZ gauge, ${\cal N}{=}4$
supersymmetry is not manifest, it should be accompanied by a field-dependent
gauge transformation to preserve the WZ gauge and the additional gauge \p{ResG}.

Here we show how to arrive at the multiplet ${\bf (1,4,3)}$ while preserving
manifest ${\cal N}{=}4$ supersymmetry.

To this end, we project all doublet $SU(2)_{PG}$ indices on the harmonics $u^{\pm i}$ using the completeness relation \p{CompL}
\bea
&& q^{+ a} = \omega u^{+ a} - l^{++} u^{- a}\,, \quad \omega = q^{+a}u^{-}_a\,, \quad
l^{++}= q^{+a}u^{+}_a\,, \label{qom} \\
&& V^{\pm \pm (ab)} = u^{+a}u^{+b}V^{\pm\pm(--)} + u^{-a}u^{-b}V^{\pm\pm(++)}
- 2u^{+(a}u^{-b)}V^{\pm\pm(+-)}\,, \nn
&& V^{\pm\pm(\pm\pm)} = V^{\pm\pm(ab)}u^{\pm}_au^\pm_b\,,\;
V^{\pm\pm(\mp\mp)} = V^{\pm\pm(ab)}u^{\mp}_au^\mp_b\,, \; V^{\pm\pm(+-)}
= V^{\pm\pm(ab)}u^{+}_au^-_b\,.
\eea
In terms of these projections, the transformation laws \p{qtraN}, \p{traN++}
and \p{traN--} read
\bea
&& \delta \omega = - \Lambda^{+-}\omega + \Lambda^{--} l^{++}\,, \quad \delta l^{++}
= \Lambda^{+-} l^{++} - \Lambda^{++}\omega\,,
\label{traNlom} \\
&& \delta V^{++(++)} = D^{++}\Lambda^{++} - 2\Lambda^{++} V^{++(+-)}
+ 2 \Lambda^{+-}V^{++(++)}\,, \nn
&& \delta V^{++(+-)} = D^{++}\Lambda^{+-} - \Lambda^{++}
- \Lambda^{++} V^{++(--)} +  \Lambda^{--}V^{++(++)}\,, \nn
&& \delta V^{++(--)} = D^{++}\Lambda^{--} - 2\Lambda^{+-}
- 2\Lambda^{+-} V^{++(--)} +  2\Lambda^{--}V^{++(+-)}\,,
\label{traN++1} \\
&& \delta V^{--(++)} = D^{--}\Lambda^{++} - 2\Lambda^{+-}
- 2\Lambda^{++} V^{--(+-)} + 2 \Lambda^{+-}V^{--(++)}\,, \nn
&& \delta V^{--(+-)} = D^{--}\Lambda^{+-} - \Lambda^{--}
- \Lambda^{++} V^{--(--)} +  \Lambda^{--}V^{--(++)}\,, \nn
&& \delta V^{--(--)} = D^{--}\Lambda^{--}
- 2\Lambda^{+-} V^{--(--)} +  2\Lambda^{--}V^{--(+-)}\,,
\label{traN--1}
\eea
while the constraint \p{NAcoq} takes the form
\bea
&& \mbox{(a)}\; D^{++}l^{++} - V^{++(+-)} l^{++} + V^{++(++)}\omega = 0\,, \nn
&& \mbox{(b)}\;
D^{++}\omega - l^{++} - V^{++(--)} l^{++} + V^{++(+-)}\omega = 0\,. \label{lomConstr}
\eea

Assuming that $<\omega> \neq 0$, we observe from the transformation law \p{traNlom}
that one can choose the following manifestly ${\cal N}{=}4$ supersymmetric gauge
\be
\omega = 1\,, \quad l^{++} = 0\, \quad \Rightarrow \quad q^{+a} = u^{+a}\,. \label{omlgauge}
\ee
In this gauge, the constraints \p{lomConstr} imply
\be
V^{++(++)} = V^{++(+-)} = 0\,, \quad V^{++(--)} \equiv {\cal V} \neq 0\,.
\ee
The residual gauge freedom is given by
\bea
&& \Lambda^{++} = 0\,, \;\; \Lambda^{+-} = 0\,, \quad \Lambda^{--} \neq 0\,, \label{ResLambd} \\
&& \delta_r {\cal V} = D^{++}\Lambda^{--}\,, \label{V++res} \\
&& \delta_r V^{--(++)} = 0\,, \; \delta_r V^{--(+-)} = -\Lambda^{--} + \Lambda^{--}V^{--(++)}\,, \nn
&& \delta_r V^{--(--)} = D^{--}\Lambda^{--} + 2\Lambda^{--}V^{--(+-)}\,. \label{V--res}
\eea
Here $\Lambda^{--} = \Lambda^{--}(\zeta, u)$ is the only unconstrained
residual analytic gauge parameter. The harmonic zero-curvature
condition \p{H0cond} is rewritten as
\bea
&& D^{++}V^{--(++)} = 0\,, \label{one} \\
&& D^{++}V^{--(+-)} - \left(1 + {\cal V}\right)V^{--(++)} + {\cal V} = 0\,,
\label{two} \\
&& D^{++}V^{--(--)} - D^{--}{\cal V}
- 2 \left(1 + {\cal V}\right) V^{--(+-)} = 0\,. \label{H0cond2}
\eea
These equations determine $V^{--(++)}\,, \,V^{--(+-)}$ and $V^{--(--)}$ as functions of the analytic gauge potential ${\cal V}\,$.

Thus in the supersymmetric gauge \p{omlgauge} we are left with the analytic gauge superfield
${\cal V}(\zeta, u)\,$, $\delta_r {\cal V} = D^{++}\Lambda^{--}\,$, as the basic object encompassing
the whole field content of the system consisting of the ${\bf (4,4,0)}$ multiplet and gauge
$SU(2)_{PG}$ superfield. The general action \p{GaugeL} takes the simple form
\bea
&& S_{gauge} = \int d t d^4\theta\,  {\cal L}(J)\,, \label{GaugeL1} \\
&& J = q^{+ a}\nabla^{--}q^{+}_{a} = 1 - V^{--(++)}\,, \quad D^{++}J = 0\,,\label{DefJ1}
\eea
where we took into account eq. \p{one}. Using the harmonic independence of $J$ in the central basis,
it is easy to find from \p{two} the expression of $J$ in terms of ${\cal V}$:
\be
J = \frac{1}{1 +{\cal W}}\,, \quad {\cal W}(t, \theta^i, \bar\theta_k)
\equiv \int du\, {\cal V}\left(t -2i\theta^i\bar\theta^k u^+_{(i}u^-_{k)}\,,\, \theta^iu^+_i,
\bar\theta^ku^+_k\,,\, u^\pm_l\right). \label{DefW}
\ee

To reveal the field content carried by ${\cal V}$, we should fully exploit the residual
infinite-dimensional gauge freedom \p{V++res}. The full WZ form of ${\cal V}$ is easily found to be
\be
{\cal V}(\zeta, u) = v_0(t_A) + \theta^+\psi^i(t_A) u^-_i + \bar\theta^+ \bar\psi^i(t_A) u^-_i +
3 \theta^+\bar\theta^+ A^{(ik)}(t_A)u^-_iu^-_k\,, \label{WZV1}
\ee
without any further residual gauge freedom, $\Lambda^{--}_r = 0\,$. Thus we end up with
the off-shell
${\cal N}{=}4$ supermultiplet ${\bf (1, 4, 3)}$ in the new formulation in terms of
the analytic gauge
prepotential ${\cal V}(\zeta, u)\,$. The off-shell transformation properties of
the component fields
in \p{WZV1} can be found from the transformation law
\be
\delta {\cal V} = (\epsilon^iQ_i + \bar\epsilon_i\bar Q^i){\cal V} + D^{++}\Lambda^{--}_{comp}\,,
\ee
where the first part is induced by the $N=4$ supertranslations \p{SUSYan} ($Q_i = -u^+_i
\frac{\partial}{\partial \theta^+} +\ldots\,, \; \bar Q^i = u^{+ i}
\frac{\partial}{\partial \bar\theta^+} +\ldots\,$), while the second part is the compensating
gauge transformation needed to preserve the WZ gauge \p{WZV1}:
\be
\Lambda^{--}_{comp} = \frac{1}{2}\left(\bar\epsilon^{(i}\bar\psi^{k)}+ \epsilon^{(i}\psi^{k)}\right)
u^-_iu^-_k +\left(\theta^+\bar\epsilon^i
-\bar\theta^+\epsilon^i \right)A^{(kl)}u^-_iu^-_ku^-_l\,.
\ee

The meaning of the ${\cal N}{=}4$ superfield ${\cal W}$ defined in
\p{DefW} can be also easily understood. First of all, by construction it is invariant under
the gauge transformations \p{V++res}, so one can always choose WZ form \p{WZV1} for ${\cal V}$
in \p{DefW}, i.e. the field content of ${\cal W}$ is just ${\bf (1,4,3)}\,$. Secondly,
using the analyticity of ${\cal V}$, $D^+{\cal V} = \bar D^+ {\cal V} = 0\,$, and the completeness
relation \p{CompL}, it is easy to show that
\bea
&& D^iD_i {\cal W} = -2\int du \,D^-D^+ {\cal V} = 0\,, \quad \bar D_i\bar D^i {\cal W} =
2\int du \,\bar D^-\bar D^+ {\cal V} = 0\,, \nn
&& [D^i, \bar D_i] {\cal W} = -2\int du\left(D^-\bar D^+ +\bar D^-D^+\right) {\cal V}
= 0\,.\label{OrdConstr}
\eea
These are just the constraints which define the ${\bf (1,4,3)}$ multiplet in the ordinary
${\cal N}{=}4$ superspace \cite{IKLe,IKPa}.

Thus we have shown that the most general sigma-model type action of
the ${\cal N}{=}4$ multiplet ${\bf (1, 4, 3)}$
can be reproduced from the most general $SU(2)_{PG}$ invariant action of the multiplet
${\bf (4,4,0)}$ by gauging the $SU(2)_{PG}$ symmetry using a ``topological'' gauge
${\cal N}{=}4$ supermultiplet.

For further use, let us note that, before imposing any
gauge-fixing condition, one can use the constraints \p{lomConstr}
to covariantly express the gauge superfields $V^{++(+-)}$ and $V^{++(++)}$ in terms of
$V^{++(--)}$, $\omega$ and $l^{++}$
\bea
&& V^{++(+-)} = \frac{1}{\omega}\left\{l^{++}\left[1 + V^{++(--)}\right] -D^{++}\omega \right\}, \nn
&& V^{++(++)} = \frac{1}{\omega^2}\left\{(l^{++})^2\left[1 + V^{++(--)}\right]
-D^{++}(l^{++}\,\omega)\right\}. \label{IH1}
\eea
Taking into account that the analytic superfields $(\omega -1)$ and $l^{++}$,
in view of their inhomogeneous transformation
laws \p{traNlom}, can be treated as Goldstone superfields related to the ``spontaneous breaking''
of local $SU(2)_{PG}$ symmetry down to its abelian subgroup with the analytic parameter
$\Lambda^{--}$, eqs. \p{IH1} supplies a nice example of the inverse Higgs phenomenon \cite{IHiggs}.
This phenomenon, in particular, provides a possibility to covariantly express gauge fields associated
with the coset generators of the given nonlinearly realized symmetry in terms of the Goldstone
fields and gauge fields belonging to the linear stability subgroup. Using the transformation laws
\p{traNlom} and \p{traN++1}, it is easy to check that the above ``composite'' gauge superfields
$V^{++(++)}$ and $V^{++(+-)}$ possess the correct gauge transformation properties. In the
``unitary'' gauge $\omega = 1, l^{++} = 0$ these superfields vanish as it should be. Also,
in accord with the general reasoning of ref. \cite{IHiggs}, one can construct a new gauge
connection
\be
\widetilde{V}^{++(--)} = \frac{1}{\omega^2}\left[(1 - \omega^2) + V^{++(--)} \right] \label{compV}
\ee
which has the ``would-be abelian'' $SU(2)_{PG}$ gauge transformation law
\be
\delta\widetilde{V}^{++(--)} = D^{++}\left(\frac{1}{\omega^2}\Lambda^{--} \right). \label{compVtran}
\ee
We have chosen $\widetilde{V}^{++(--)}$ in such a way that it is equal to $V^{++(--)}$
in the unitary gauge. In what follows, this gauge connection will be used to construct
an invariant Fayet-Iliopoulos (FI) term for the multiplet ${\bf (1,4,3)}\,$.

\subsection{Superconformal properties}
Now we turn to discussing the superconformal properties of the new description of the multiplet
${\bf (1, 4, 3)}$ and some immediate applications thereof. We shall need to know the realization
of the most general ${\cal N}{=}4, d=1$ superconformal group $D(2,1;\alpha)$ in the analytic
basis of ${\cal N}{=}4$ superspace \cite{IL}. Since the whole set of the $D(2,1;\alpha)$
transformations is contained in the closure of Poincar\'e supersymmetry \p{SUSYan} and conformal supersymmetry, it is sufficient to explicitly give only the transformations of the
latter:
\bea
&&\delta' \theta^+ = \eta^+\,t_A + 2i(1{+}\alpha)
\eta^-(\theta^+\bar\theta^+)\,, \quad
\delta' \bar\theta^+ = \bar\eta^+\,t_A +
2i(1{+}\alpha)\bar\eta^-(\theta^+\bar\theta^+)\,,
\label{trthet+} \\[6pt]
&&\delta' u^+_i = \Lambda^{++}_{sc} u^-_i\,, \; \delta' u^-_i = 0\,, \;
\Lambda^{++}_{sc} = -2i\alpha(\eta^+ \bar\theta^+ {-} \bar\eta^+ \theta^+ )
\equiv D^{++}\Lambda_{sc}\,, \nn [6pt]
&& \Lambda_{sc} = -2i\alpha(\eta^- \bar\theta^+ {-} \bar\eta^- \theta^+ )\,,
\quad (D^{++})^2\Lambda_{sc} = 0\,,
\label{defLambda} \\[6pt]
&& \delta' \theta^- = \eta^- t_A + 2i \eta^-[\,(1{+}\alpha) \theta^+\bar\theta^-
+ \theta^-\bar\theta^+\,]
+ 2i\alpha \,\bar\eta^-\theta^-\theta^+
-2i(1{+}\alpha) \eta^+ \theta^-\bar\theta^-\,, \label{trthet-} \\[6pt]
&& \delta' t_A = -2i t_A\,(\eta^-\bar\theta^+  {-} \bar\eta^-\theta^+)\,,
\quad \delta' \bar\theta^- = \widetilde{\delta' \theta^-}\,.\label{t-trans}
\eea
Here, $\eta^\pm = \eta^iu^\pm_i, \bar\eta^\pm = \bar\eta^iu^\pm_i$ and $\eta^i,   \bar\eta^i$ are
the corresponding Grassmann parameters and the involution $\widetilde{\;\;\;}$ is defined in \p{Tilde}. We also need the transformation properties
of the harmonic derivatives, the ${\bf (4,4,0)}$ superfield $q^{+ a}\,$,
gauge potentials $V^{\pm\pm}$
and the measures of integration over the full and
analytic harmonic superspaces \p{measures}
\bea
&& \delta' D^{++} = -\Lambda^{++}_{sc} D^0\,, \quad \delta' D^{--} = -(D^{--}\Lambda^{++}_{sc})\, D^{--}\,, \;
\delta' D^0 = 0\,, \label{DTrans} \\[6pt]
&& \delta' q^{+ a} = \Lambda_{sc}\, q^{+ a}\,, \quad \delta' V^{++} = 0\,, \quad
\delta' V^{--}= -(D^{--}\Lambda^{++}_{sc})\, V^{--}\,, \label{FieldTrans} \\[6pt]
&& \delta' \mu_A^{(-2)} =
\left(\partial_A\delta' t_A + \partial^{--}\Lambda^{++}_{sc} -
\partial_{\theta^+}\delta' \theta^+ -
\partial_{\bar\theta^+}\delta' \bar\theta^+\right)\mu_A^{(-2)} = 0\,,
\label{muA}\\[6pt]
&& \delta' \mu_H =
\left(\partial_A\delta' t_A + \partial^{--}\Lambda^{++}_{sc} -
\partial_{\theta^+}\delta' \theta^+ -
\partial_{\bar\theta^+}\delta' \bar\theta^+
- \partial_{\theta^-}\delta' \theta^- -
\partial_{\bar\theta^-}\delta' \bar\theta^-\right)\mu_H \nonumber\\[6pt]
&& \qquad \; =\, 2i\left[ (1{-}\alpha)(\eta^-\bar\theta^+ -\bar\eta^-\theta^+) -
(1{+}\alpha)(\eta^+\bar\theta^- -\bar\eta^+\theta^-)\right]\mu_H\,.
\label{muH}
\eea
The integration measures are evidently invariant under the ${\cal N}{=}4$ Poincar\'e supersymmetry \p{SUSYan}.
Note that for the special values of the parameter
$\alpha $,
\be
\mbox{(a)}\;\; \alpha = -1\,, \qquad \mbox{(b)}\;\;\alpha = 0\,, \label{spec}
\ee
the supergroup $D(2,1;\alpha)$ is reduced to two different $PSU(1,1|2)$ supergroups, such that
one of the two commuting $R$-symmetry subgroups $SU(2) \subset D(2,1;\alpha)$ is identified with $SU(2) \subset PSU(1,1|2)$, while the other decouples and
acts as outer automorphisms of this $PSU(1,1|2)\,$. In particular, as follows from
\p{defLambda}, the $PSU(1,1|2)$ supergroup corresponding to the choice $\alpha = 0$ does not affect
harmonic variables at all.

It is easy to check superconformal covariance of the constraints \p{coq}, \p{NAcoq} and the
harmonic zero-curvature conditions \p{Hzc}, \p{H0cond} at any $\alpha$. The quantity $J$ defined
in \p{DefJ1} is transformed as
\be
\delta' J = \left(2\Lambda_{sc} - D^{--}\Lambda^{++}_{sc}\right) J \equiv \Lambda_0\, J\,,
\quad D^{++}\Lambda_0 = 0\,. \label{Jconf}
\ee
The superconformal properties of the basic quantities in the gauge \p{omlgauge} can be easily found
from the condition of preserving this gauge (which fixes the relevant compensating
gauge transformations) and the transformation property of the harmonic measure $du$
in the central basis
\be
\delta' du = du\,D^{--}\Lambda^{++}_{sc}\,.
\ee
We obtain
\be
\delta' {\cal V} = -2\Lambda_{sc}\left(1 + {\cal V} \right), \quad \delta' V^{--(++)} =
-\Lambda_0 \left[1 - V^{--(++)} \right], \quad \delta' {\cal W} = -\Lambda_0\,
\left(1 + {\cal W}\right)\,. \label{GaugeConf}
\ee
Taking into account that in this gauge $J = 1 - V^{--(++)}$, we see that \p{GaugeConf}
agrees with \p{Jconf}. Also, it is convenient to define ${\cal U} = 1+ {\cal W}$, so that
\be
J = \frac{1}{{\cal U}}\,, \quad \delta' J = \Lambda_0\, J\,, \quad \delta' {\cal U}
= -\Lambda_0\, {\cal U}\,. \label{Utransf}
\ee
The object ${\cal U}$ satisfies the same constraints \p{OrdConstr} as ${\cal W}$, but has simpler
transformation properties.

Let us recall the form of superconformally invariant actions of the multiplet ${\bf (1,4,3)}$.
Within the above gauging procedure, for all $\alpha$ except the special value $\alpha =0$,
they are uniquely defined by the corresponding actions of
the multiplet ${\bf (4,4,0)}$. The latter are formulated in terms of
$q^{+ a}q^-_a \sim q^{ia}q_{ia}$ \cite{IL}. To obtain the invariant action of thee multiplet
${\bf (1, 4, 3)}$, one just should make the replacement
$q^{+ a}q^-_a \rightarrow J = q^{+ a}\hat q^-_a$ in the corresponding ${\bf (4,4,0)}$ action.
For $\alpha \neq 0, -1$ such subclass of the general sigma-model action \p{qact1} is given by
\be
S^\alpha_{(sc)} = \gamma \int dt d^4\theta\, J{\,}^{\frac{1}{\alpha}} =
\gamma \int dt d^4\theta\, {\cal U}{\,}^{-\frac{1}{\alpha}}\,, \label{scinv}
\ee
where $\gamma$ is a normalization constant. In particular, the free $q^+$ action \p{Freeq}
is invariant under the supergroup $D(2,1;\alpha=1\,) \sim OSp(4^\star\vert 2)$, and the associated superconformal ${\bf(1,4,3)}$ action is just
\be
S^{\alpha=1}_{(sc)} = \gamma\int dt d^4\theta\, J =\gamma\int dt d^4\theta\, {\cal U}{\,}^{-1}\,.
\label{alpha1}
\ee
This action contains a non-trivial self-interaction, despite the fact that it was obtained
by gauging the free $q^{+ a}$ action. It is the only one which admits a local representation
as an integral over the analytic superspace. This is just the example we have
considered in \cite{I}. On the other hand, the choice of $\alpha = -1/2$ yields
the free ${\bf(1, 4, 3)}$ action
\be
S^{\alpha=-1/2}_{(sc)} = \gamma\int dtd^4\theta\, {\cal U}{\,}^{2}\,, \label{freeU}
\ee
though it is obtained by gauging a non-trivial sigma-model $q^{+}$ action, with the superfield
Lagrangian ${\cal L} \sim (q^{ia}q_{ia})^{-2}$ in \p{qact1}\footnote{In ref. \cite{root}
there is an erroneous statement that all superconformally invariant actions of the reduced
multiplets are generated from the free  $q^{+ a}$ action.}.

The cases of  $\alpha = -1$ and $\alpha =0$ at which $D(2,1;\alpha)$ degenerates
into the $PSU(1,1|2)$ supergroups (times outer $SU(2)$ automorphisms) require a separate consideration.

At $\alpha = -1$ the action \p{scinv} is still invariant, but now
the Lagrangian is just ${\cal U} = 1 + {\cal W}$ and the action identically
vanishes as a consequence of the constraints \p{OrdConstr}. The meaningful action
in this case reads \cite{IKLe,IKL0}
\be
S^{\alpha=-1}_{(sc)} \sim \int dtd^4\theta\, {\cal U} \log {\cal U}\,,\label{-1act}
\ee
which is invariant under $PSU(1,1|2)$ (and the extra $SU(2)$ automorphisms), up to a total derivative in the integrand.

In the case of $\alpha = 0$ the situation is even more subtle. As seen from \p{defLambda},
in this case the harmonic variables are inert under the corresponding $PSU(1,1|2)$ \footnote{They are still transformed under the extra automorphisms $SU(2)$ acting on the indices $i, k\,$.} and the superfields ${\cal V}\,$, ${\cal U}$
are transformed with zero conformal weight. On the other hand, the integration measure $\mu_H$
is still not invariant,
\be
\delta' \mu_H =\, 2i\left(\bar\eta^i\theta_i - \eta^k\bar\theta_k \right)\mu_H \,.
\ee
The only way to construct the invariant action in this case is to modify the transformation law of
the analytic prepotential ${\cal V}$ under the conformal supersymmetry:
\be
\delta'_{mod} {\cal V} = 4i\left(\bar\eta^-\theta^+ - \eta^-\bar\theta^+\right) \label{modTrans}
\ee
(actually, the coefficient in the r.h.s. can be an arbitrary non-zero real number;
it was chosen in this way just for further convenience, using
the freedom of rescaling ${\cal V}\,$).
Respectively, the superfield ${\cal U}$ is now transformed as
\be
\delta'_{mod} {\cal U} = - 2i\left(\bar\eta^i\theta_i - \eta^k\bar\theta_k \right). \label{modTrans1}
\ee
The extra terms in some other $PSU(1,1|2)$ transformations can be found by taking the Lie
brackets of \p{modTrans}, \p{modTrans1} with Poincar\'e supersymmetry (in fact, when acting  on ${\cal V}\,$, the symmetry $PSU(1,1|2)$ closes modulo some particular gauge transformation). The invariant action for the case $\alpha = 0$ is then as follows
\be
S^{\alpha=0}_{(sc)} \sim \int dtd^4\theta\, e{\,}^{\cal U}\,. \label{0act}
\ee

The modified $\alpha = 0$ superconformal transformation \p{modTrans}
cannot be obtained from any modification of the $\alpha = 0$
transformations of $q^{+ a}$ and/or  $V^{++}{}^{(ab)}$ before fixing the unitary gauge
with respect to the local $SU(2)_{PG}$. This becomes possible within the alternative
gauging of the ${\bf (4,4,0)}$ system considered in the next Section.

Let us now discuss the issue of superconformally invariant potentials of the multiplet
${\bf (1,4,3)}$ in the description through the analytic prepotential ${\cal V}\,$.
It is easy to see that there exist no $SU(2)_{PG}$ invariant WZ-type $q^{+a}$
Lagrangians among those in \p{WZq}. Thus it seems impossible to generate the potential terms for the multiplet ${\bf (1,4,3)}$
by $SU(2)_{PG}$ gauging of any $q^{+ a}$ action. Nevertheless, such terms can be constructed
with the help of the gauge superfield  $V^{++(ab)}\,$. Prior to imposing any $SU(2)_{PG}$ gauge,
one can define the gauge invariant FI term
\be
S_{FI} = \int du d\zeta^{(-2)}\, c^{+2}\, \widetilde{V}^{++(--)} = \int du d\zeta^{(-2)}\, c^{+2}\,
{\cal V}\, \quad c^{+2} = c^{ik}u^+_iu^+_k\,, \;[c] = cm^{-1}\,,\label{FI1}
\ee
where $\widetilde{V}^{++(--)}$ was defined in \p{compV}, \p{compVtran}. This action is invariant
under \p{compVtran} and \p{V++res} thanks to the condition $D^{++}c^{++} = 0$,
and in \p{FI1} we made use of the property that $\widetilde{V}^{++(--)} = {\cal V}$ in the unitary
gauge $\omega = 1, l^{++} = 0\,$.
In the WZ gauge \p{WZ} the component Lagrangian following from  \p{FI1}
is $\;\;\propto\; c^{ik}A_{ik}\,$. When
\p{FI1} is added to some non-trivial action \p{GaugeL1}, eliminating the auxiliary
field $A_{ik}$ gives rise to a non-trivial potential of the physical bosonic field $v_0$
(plus the appropriate Yukawa-type fermionic terms) \cite{IKLe,IKPa}.

Inspecting the superconformal properties of \p{FI1}, one can check that it is
superconformally ($PSU(1,1|2)$) invariant only at $\alpha = 0\,$ (the shift \p{modTrans} is linear
in the analytic coordinates $\theta^+, \bar\theta^+$ and so does not affect \p{FI1}).
Thus one possibility to construct the superconformally invariant action
of the multiplet ${\bf (1,4,3)}$ with a non-trivial scalar potential is to sum up \p{FI1} with
\p{0act},
\be
\tilde{S}{\,}^{\alpha = 0}_{(sc)} = \gamma \int dtd^4\theta\, e{\,}^{\cal U} +
\int du d\zeta^{(-2)}\, c^{+2}\, {\cal V}\,\,. \label{FI+0}
\ee
After elimination of $A_{ik}$ there comes out the conformal potential $\sim e^{-(1+v_0)}$ with
the strength $\sim c^2 = c^{ik}c_{ik}\,$ (see below). The extra automorphisms $SU(2)$ acting on the indices $i, k$ is obviously broken down to some $U(1)$ due to the presence of the constant triplet $c^{ik}$ in \p{FI1}, \p{FI+0}.

The only alternative mechanism of generating superconformally invariant potential term for
the multiplet ${\bf (1,4,3)}$ which was known to date \cite{IKLe} requires $\alpha = -1$.
Though \p{FI1} is not superconformally invariant in this case and so cannot be used,
one can modify the constraints \p{OrdConstr}
by inserting two arbitrary constants (one complex and one real) in their r.h.s.
They form a constant isotriplet with respect to the second $R$-symmetry $SU(2)$
subgroup (the one which provides outer automorphisms of $PSU(1,1|2)$ corresponding to the choice $\alpha = -1\,$).
Exploiting this broken $SU(2)$ symmetry, one can choose the frame where
\be
(D)^2 \,\widetilde{{\cal U}} = (\bar D)^2\, \widetilde{{\cal U}} = 0\,,
\quad [D^i,\bar D_i]\, \widetilde{{\cal U}} = f\,, \quad f = f^*\,.
\ee
With $\widetilde{{\cal U}}$ transforming as in \p{Utransf}, at $\alpha = -1$
the expressions in the l.h.s.
of these constraints can be checked to be scalars of zero conformal weight, so one can equate
them to some non-zero constants without contradiction with the superconformal $PSU(1,1|2)$ symmetry.
The substitution of $\widetilde{{\cal U}}$ into \p{-1act} for ${\cal U}$ once again
yields the conformal potential for $v_0$ in the component action (with strength $\sim f^2\,$).

It is interesting to see how the modified constraints emerge within the analytic prepotential description of the multiplet
${\bf (1,4,3)}$. The superfield $\widetilde{{\cal U}}$ is related to ${\cal U}
= 1 + \int du\,{\cal V}$ in the following way
\be
\widetilde{{\cal U}} = {\cal U} + \frac{1}{2}\,f\,\bar\theta^i\theta_i = {\cal U} +
\frac{1}{2}\,f\left(\bar\theta^+\theta^- - \bar\theta^-\theta^+ \right). \label{tildeU}
\ee
This superfield has precisely the same transformation properties with respect to
$PSU(1,1|2)$ corresponding to $\alpha =-1\,$ as the superfield ${\cal U}$, i.e. $\delta \widetilde{{\cal U}} =
-\Lambda_{0}^{(\alpha=-1)}\,\widetilde{{\cal U}}\,$,  provided that the analytic
prepotential ${\cal V}\,$ possesses the following modified transformation rules
with respect to the Poincar\'e and conformal supersymmetries
\be
\widetilde{\delta} {\cal V} = -2\Lambda_{sc}^{(\alpha=-1)} (1+ {\cal V}) +
f\left[(\epsilon^- +t_A\eta^-)\bar\theta^+ + (\bar\epsilon{}^-
+t_A\bar\eta^-)\theta^+ \right]. \label{modTransI}
\ee
It is easy to check that the closure of the modified transformations coincides
with the original one (i.e. for $f=0$)
modulo some special gauge transformation of the form \p{V++res}. The latter
does not make contribution to
${\cal W} = \int du {\cal V}$ owing to the $u$-integral. Note that the second term
in \p{tildeU} cannot be re-absorbed into ${\cal V}$ because it contains non-analytic Grassmann
coordinates.

It is curious that the prepotential realization of the ${\bf (1,4,3)}$ multiplet suggests one more
mechanism of generating conformal potential for the bosonic field $v_0\,$. Again, it only works  for the
$PSU(1,1|2)$ case with $\alpha = -1\,$. It is non-minimal, because it uses a superconformal
coupling to the extra off-shell multiplet ${\bf (0,4,4)}\,$. The latter contains no
physical bosons at all and comprises 4 fermionic fields and 4 bosonic auxiliary fields.
It is described by the fermionic analog of $q^{+ a}$, the superfield $\Psi^{+ m}$,
$\widetilde{(\Psi^{+ m})} = \Psi_m^+\,$, subjected to the constraint \cite{IL}
\be
D^{++}\Psi^{+ m} = 0\,\; \Rightarrow \; \Psi^{+ m} =
\psi^{im}u^+_i + \theta^+ a^m + \bar\theta^+ \bar{a}^m
+ 2i\theta^+\bar\theta^+ \partial_t\psi^{im}u^-_i\,.  \label{PsiConstr}
\ee
With respect to the doublet index $m$ ($m = 1,2$), it is transformed by some extra $SU(2)$ which
commutes with ${\cal N}{=}4$ and so is an analog of $SU(2)_{PG}$ (it does not necessarily coincide
with the latter). The requirement of superconformal covariance of the constraint \p{PsiConstr} uniquely fixes the superconformal
$D(2,1;\alpha)$ transformation
rule of $\Psi^{+ m}$, for any $\alpha$, as follows
\be
\delta' \Psi^{+ m} = \Lambda_{sc} \Psi^{+ m}\,.\label{traNPsi1}
\ee

The free action of $\Psi^{+ m}$,
\be
S^{\psi}_{free} = \int du d\zeta^{(-2)}\Psi^{+ m}\Psi^{+}_{m}\,, \label{Freepsi}
\ee
is obviously not invariant under $D(2,1;\alpha)\,$. However, recalling the transformation law
\p{GaugeConf}, we observe that this action can be easily promoted to a superconformal
invariant by coupling $\Psi^{+ m}$ to the ${\bf (1,4,3)}$ multiplet
\be
S^{\psi}_{(sc)} =\int du d\zeta^{(-2)}( 1 + {\cal V})\Psi^{+ m}\Psi^{+}_{ m}\,. \label{Confpsi}
\ee
This action is superconformal at any $\alpha $, and it also respects the gauge invariance
\p{V++res} as a consequence of the constraint \p{PsiConstr}. However, a simple analysis
shows that in components it yields only a bilinear term in the auxiliary fields $a^m, \bar a^m$
and therefore cannot produce a non-trivial potential of the field $v_0$. To get such a potential,
one needs to add a FI-type term to \p{Confpsi}
\be
S^{\psi}_{FI} = \int du d\zeta^{(-2)} \left( \theta^+ \xi_m\Psi^{+ m}
+ \bar\theta^+ \bar\xi^m\Psi^+_m \right), \label{FIpsi}
\ee
where $\xi_m, \bar\xi^m$ is a constant doublet which breaks the extra $SU(2)$
acting on the indices $m\,$. Using the transformation
properties \p{trthet+}, \p{muA} and \p{traNPsi1}, as well as the constraint \p{PsiConstr},
it is easy to show that \p{FIpsi}
at $\alpha = -1$ is invariant under both Poincar\'e and conformal supersymmetries,
up to a total derivative
in the integrand. After passing to components, this action produces terms which
are linear in the auxiliary
fields $a^m, \bar a^m\,$. Eliminating these fields in the total
$D(2,1; \alpha=-1)$ invariant action
\be
S_{tot} = S^{\alpha=-1}_{(sc)} + S^{\psi}_{(sc)} + S^{\psi}_{FI}\,, \label{totalpsi}
\ee
one again reproduces the conformal potential for $v_0\,$. The price for this is the enlargement
of the physical fermionic sector of the model from $4$ to $8$ fields, still with the presence of only one physical bosonic field. Taking into account that the ${\cal N}{=}4$ multiplets ${\bf (1,4,3)}$
and ${\bf (0,4,4)}$ can be joined into the off-shell ${\cal N}{=}8$ multiplet
${\bf (1, 8, 7)}$ \cite{BIKL}, one can expect a hidden ${\cal N}{=}8$ supersymmetry
in this combined system.

Note that the two mechanisms of obtaining superconformally invariant potential terms at
$\alpha = -1$ described above cannot coexist since the action \p{Confpsi} is not invariant under the
modified superconformal transformations \p{modTransI}.

\subsection{Examples of component actions}
Let us give a few examples of bosonic component actions. We shall need the form of the bosonic
sector of the superfield $J$ and ${\cal U} = 1 + {\cal W}$ defined in \p{DefJ1} and \p{DefW}.
Passing to the central basis in the prepotential ${\cal V}(\zeta, u)$, we find from \p{DefW}
\bea
{\cal U} &=& 1 + {\cal W} = (1 +v_0) + [\theta^-\bar\theta^-A^{++} -(\theta^-\bar\theta^+
+ \theta^+\bar\theta^-)A^{+-} + \theta^+\bar\theta^+ A^{--}] \nn
&& + \,\theta^+\bar\theta^+\theta^-\bar\theta^-\,\partial_t^2 v_0\,, \label{expand}
\eea
where all component fields are functions of $t$ and we still used the harmonic projections of
the Grassmann coordinates (in fact, the harmonic dependence in \p{expand} is fake, which immediately
follows from the easily checkable relation $\partial^{++}{\cal U} = 0\,$). The corresponding
expression for the superfield $J = {\cal U}^{-1}$ is
\bea
J &=& \frac{1}{1 + v_0}\,\{1 - \frac{1}{1 + v_0}[\theta^-\bar\theta^-A^{++} -(\theta^-\bar\theta^+
+ \theta^+\bar\theta^-)A^{+-} + \theta^+\bar\theta^+ A^{--}] \nn
&& -\,\frac{1}{1 + v_0}\theta^+\bar\theta^+\theta^-\bar\theta^-
\,[\partial_t^2 v_0 - \frac{1}{1+v_0} A^{ik}A_{ik}]\}\,.
\eea
Using these explicit expressions, we find, in particular,
\bea
&& S^{\alpha =1}_{(sc)} \;\Rightarrow \;S^{\alpha =1}_{bos} \sim \int dt\,[\,(\partial_t \rho)^2
- \frac{1}{8} \rho{\,}^6 (A^{ik}A_{ik})\,]\,, \quad \rho = (1 + v_0)^{-1/2}\,, \label{bos1} \\
&& S^{\alpha =-1}_{(sc)} \;\Rightarrow \; S^{\alpha =-1}_{bos}
\sim \int dt\,[\,(\partial_t \rho)^2
- \frac{1}{8\rho{\,}^2} (A^{ik}A_{ik})\,], \quad \rho = \sqrt{(1 + v_0)}\,, \label{bos-1} \\
&& S^{\alpha =0}_{(sc)} \;\Rightarrow \; S^{\alpha =0}_{bos}
\sim \int dt\,[\,(\partial_t \rho)^2
- \frac{1}{8}\,\rho{\,}^2\, (A^{ik}A_{ik})\,],\quad \rho = 2 e{\,}^{\frac{1}{2}(1 + v_0)}\,,
\label{bos0}
\eea
where the superfield actions were defined in \p{alpha1}, \p{-1act} and \p{0act}. For $\widetilde{{\cal U}} = {\cal U} + \frac{1}{2}\,f\,(\bar\theta^+\theta^-
- \bar\theta^-\theta^+)$
there appears the additional (conformal) potential term in \p{bos-1}
$$
-\frac{1}{16}\int dt\,\frac{f{\,}^2}{\rho{\,}^2}\,.
$$

The FI term \p{FI1} yields
\be
S_{FI} \; \Rightarrow \; S_{FI}^{bos} = i\,\int dt \,c^{ik}A_{ik}\,.
\ee
After eliminating $A^{ik}$ from the total action \p{FI+0} (with $\gamma =-1$ for simplicity), one obtains
\be
\tilde{S}_{(sc)}^{\alpha = 0} \; \Rightarrow \; \tilde{S}_{bos}^{\alpha=0}
= \int dt\,[\,(\partial_t \rho)^2
- \frac{2 c^2}{\rho{\,}^2}\,]\,.
\ee

The sum of actions \p{Confpsi} and \p{FIpsi} gives rise to the following bosonic contribution
\be
S^{\psi}_{(sc)} + S^{\psi}_{FI} \; \Rightarrow \; \int dt \,[\,2(1+v_0)\,a^m\bar a_m
+ (\,\xi_m\bar a^m - \bar\xi^m a_m\,)\,]\,,
\ee
which, upon eliminating the auxiliary fields $a^m, \bar a_m\,$, again adds the conformal
potential to the action \p{bos-1} in the total action \p{totalpsi}:
\be
S^{\alpha =-1}_{bos} \; \Rightarrow \; S{\,}^{\alpha =-1}_{bos} - \frac{1}{2}\int dt \,
\frac{\xi_m\bar\xi^m}{\rho{\,}^2}\,.
\ee

\subsection{``Mirror'' (1, 4, 3) multiplet}
To close this Section, we make a few comments on the description of the ``mirror''
${\bf (1, 4, 3)}$ multiplet in the considered setting. The basic difference between
this multiplet \cite{IKPa} and the one discussed above is that its three auxiliary fields form
a triplet with respect to
the second (hidden) $SU(2)$ automorphism group of ${\cal N}{=}4$, $d=1$ Poincar\'e superalgebra.
They are singlets
with respect to the manifest automorphism $SU(2)$ acting on the doublet indices $i$
of harmonics and
Grassmann coordinates. One could consider an alternative ${\cal N}{=}4$, $d=1$
harmonic superspace, with just
this second $SU(2)$ being harmonized. In this superspace the ``mirror''
${\bf (1, 4, 3)}$ multiplet
is described in the same way as the multiplet we dealt with here, the only difference
being in the
$D(2,1;\alpha)$ superconformal properties, such that the special cases $\alpha = 0$ and
$\alpha = -1$ switch with respect to each other (the formal coincidence with the description in the ${\cal N}=4, d=1$ harmonic superspace considered here can be restored by passing to the parameter $\beta = -(\alpha +1)\,$). On the other hand, in the framework
of the harmonic superspace
considered here this alternative ${\bf (1,4,3)}$ multiplet is described by
a general superfield $\Omega$ subjected to the following constraints \cite{I}:
\be
\mbox{(a)} \;\;D^+ \bar D^+ \Omega = 0\,, \quad \mbox{(b)}\;\;
D^{++}\Omega = 0\,.
\lb{Omega1}
\ee
In the analytic basis, these constraints imply
\be
\Omega = \Sigma(\zeta, u) + i\left[\theta^- \Psi^+(\zeta, u)
+ \bar\theta^-\bar\Psi^+(\zeta, u)\right], \lb{Omega2}
\ee
\be
\mbox{(a)}\;\;D^{++}\Psi^+ = D^{++}\bar\Psi^+ = 0\,, \quad \mbox{(b)}\;\;
D^{++}\Sigma +
i\left(\theta^+ \Psi^+ + \bar\theta^+\bar\Psi^+ \right) = 0\,. \lb{Omega3}
\ee
The general solution of \p{Omega3} is
\bea
&& \Psi^+ = \psi^iu^+_i + \theta^+ s + \bar\theta^+ r
+ 2i \theta^+\bar\theta ^+\partial_t \psi^i u^-_i\,, \nn
&& \bar\Psi^+ = -\bar\psi^i u^+_i - \theta^+ \bar r + \bar\theta^+ \bar s -
2i \theta^+\bar\theta^+ \partial_t \bar\psi^i u^-_i\,, \label{143Psi}\\
&& \Sigma = \sigma - i\theta^+ \psi^iu^-_i
+ i\bar\theta^+ \bar\psi^i u^-_i\,,\label{143}
\eea
where
\be
\mbox{Re}\,r = \partial_t \sigma\,. \lb{add}
\ee
The independent fields $\sigma(t), \psi^i(t), s(t), \mbox{Im}\, r(t)$
constitute the alternative off-shell ${\bf (1,4,3)}$ multiplet.

The superconformal properties of the superfields $\Sigma$, $\Psi, \bar\Psi$
can be easily defined and the relevant actions can be constructed
analogously to those presented above. Once again, the superconformally
invariant potential terms can be constructed only for $\alpha = 0$ and $\alpha = -1\,$.
At $\alpha = 0\,$, the expression in the l.h.s. of (\ref{Omega1}a) has the conformal
weight zero, so one can consider the more general condition
\be
D^+ \bar D^+ \Omega = \tilde{c}^{++}\,, \quad \tilde{c}^{++} = \tilde{c}^{ik}u^+_iu^+_k\,,
D^{++}\tilde{c}^{++} = 0\,. \lb{Omega1mod}
\ee
The constants $\tilde{c}^{ik}$ have the dimension of mass and, via the constraint (\ref{Omega1}b),
properly modify \p{Omega2} - \p{143}. After substitution of the modified superfields into
the sigma-model type superconformal action of $\Omega$, one obtains the conformal potential
for $\sigma_0\,$, in the same way as for $v_0$ in the case $\alpha = -1$. In the latter case,
the mechanism of activating the superconformal potential for the ``mirror'' multiplet resembles
the $\alpha =0$ one for the multiplet ${\bf (1, 4, 3)}$ of the first kind: it consists
in adding the proper FI-type term to the corresponding invariant sigma-model action
\be
S_{FI}{}' = \int dud\zeta^{(-2)}\left(g_1 \theta^+\Psi^+ + g_2 \theta^+ \bar\Psi^+ + \mbox{c.c.}
\right).
\ee
This action can be shown to be $D(2,1;\alpha)$ invariant only for $\alpha = -1\,$. The
non-minimal mechanism of obtaining the superconformally invariant potentials exists in this case too,
now for the choice $\alpha =0\,$. Also, it is easy to couple the two kinds
of ${\bf (1,4,3)}$ multiplets to each
other through an interaction similar to \p{Confpsi}
\be
S_{I-II} \sim \int dud\zeta^{(-2)} (1 + {\cal V})\Psi^+\bar\Psi^+\,,
\ee
where $\Psi, \bar\Psi$ satisfy the constraints (\ref{Omega3}a) corresponding to the choice
\p{Omega1} (i.e. for $\tilde{c}^{ik} = 0$ in \p{Omega1mod}). In the future we hope to come
back to a more
detailed analysis of these models and their possible implications
in such long-standing problems as constructing ${\cal N}{=}4$ extensions
of the Calogero and Calogero-Moser integrable systems \cite{GT}.

\setcounter{equation}{0}
\section{Gauging shift isometries}
The multiplet ${\bf (1,4,3)}$ can equally be reproduced by gauging three mutually commuting
shift isometries (\ref{Shift1}b).

After promoting $\tilde\lambda^a_{\;\;b}$ in  (\ref{Shift1}b) to
$\tilde\Lambda^a_{\;\;b}(\zeta, u)$,
\be
\delta q^{+ a} = \tilde{\Lambda}^a_{\;\;b} u^{+ b}\,, \quad \tilde{\Lambda}^a_{\;\;a} =0\,,
\label{ShTr}
\ee
the constraint \p{coq} should be covariantized by introducing three abelian
analytic gauge connections
$V^{++(ab)}$
\be
D^{++}q^{+ a} + V^{++(ab)}u^+_b = 0\,, \quad \delta V^{++(ab)}
= D^{++}\tilde{\Lambda}^{(ab)}\,. \label{constrSh}
\ee
Like in the $SU(2)_{PG}$ case, one can introduce non-analytic
gauge connection $V^{--(ab)}$,
\be
D^{++}V^{--(ab)} - D^{--}V^{++(ab)} = 0\,, \quad \delta V^{--(ab)} = D^{--}\tilde{\Lambda}^{(ab)}\,,
\label{HarmFl}
\ee
and define the non-analytic superfield $\hat{q}^{- a}$ as
\bea
&& \hat{q}^{- a} = \nabla^{--}q^{+ a} = D^{--}q^{+ a} + V^{-- (ab)}u^+_b\,, \quad
\delta \hat{q}^{- a} = \tilde{\Lambda}^a_{\;\;b} u^{- b}\,; \label{defq-1} \\
&& \nabla^{++}\hat{q}^{- a} = D^{++}\hat{q}^{- a} + V^{++(ab)}u^-_b = q^{+ a}\,, \;
\nabla^{--}\hat{q}^{- a} = D^{--}\hat{q}^{- a} + V^{--(ab)}u^-_b = 0\,. \label{Prop}
\eea
Note that $\delta \nabla^{--}\hat{q}^{- a} =0\,$, and the last relation  in \p{Prop}
follows from the equation $\nabla^{++}\nabla^{--}\hat{q}^{- a}
= D^{++}\nabla^{--}\hat{q}^{- a} = 0\,$, which can be easily proved using the constraint
in \p{constrSh} and the harmonic flatness condition in \p{HarmFl}.

Passing to the harmonic projections $\omega, l^{++}$ and $V^{++(\pm\pm)}, V^{++(+-)}$
by the same relations as in the previous Section, we observe that
\be
\delta \omega = -\tilde\Lambda^{+-}\,, \quad \delta l^{++} =
-\tilde\Lambda^{++}\,, \label{tildeTran}
\ee
where the involved analytic parameters are the proper projections of
$\tilde{\Lambda}^{(ab)}$. Eqs. \p{tildeTran} suggest the choice
of the manifestly supersymmetric
unitary gauge in this case as
\be
\omega = l^{++} = 0\,. \label{Unit2}
\ee
In this gauge, as follows from \p{constrSh},
\be
V^{++(++)} = V^{++(+-)} = 0\,\label{Vconstr1}
\ee
and the only residual gauge symmetry is
\be
\delta V^{++(--)} = D^{++}\tilde\Lambda^{--}\,, \quad
\tilde\Lambda^{--} = \tilde\Lambda^{--}(\zeta, u)\,.
\ee
The harmonic flatness condition is reduced to the set
\bea
&& D^{++}V^{--(++)} = 0\,, \label{one1} \\
&& D^{++}V^{--(+-)} - V^{--(++)} + {\cal V} = 0\,, \quad {\cal V} \equiv V^{++(--)}\,,
\label{two1} \\
&& D^{++}V^{--(--)} - D^{--}{\cal V} - 2 V^{--(+-)} = 0\,, \label{H0cond3}
\eea
which is just the abelian version of \p{one} - \p{H0cond2}. We see that,
like in the case of $SU(2)_{PG}$ gauging,
the basic object encoding the irreducible field content in the unitary gauge is the analytic
superfield ${\cal V} = V^{++(--)}$ with the gauge transformation law
$\delta {\cal V} = D^{++}\Lambda^{--}$ which allows one to choose the WZ gauge
\p{WZV1} with the ${\bf (1,4,3)}$ off-shell field content. From eqs. \p{one1}, \p{two1}
we deduce the expression for $V^{--(++)}$ in terms of ${\cal V}$
which coincides with the one given in eq. \p{DefW}
\be
V^{--(++)}\equiv {\cal W} = \int du \,{\cal V}\,. \label{CalW}
\ee

The remaining projections $V^{--(+-)}$ and $V^{--(--)}$
can be also expressed through ${\cal V}$ from
\p{one1} - \p{H0cond3}; like in the $SU(2)_{PG}$ case, they seem to be of no need
for constructing the corresponding invariant actions. Indeed, the superfield $V^{--(++)}
= {\cal W}$
in eq. \p{CalW} by construction satisfies the constraints \p{OrdConstr}
defining the off-shell multiplet
${\bf (1, 4, 3)}$ in the ordinary ${\cal N}{=}4$ superspace. The most general sigma-model type
off-shell action of this multiplet is given by an expression similar to \p{GaugeL1}
\be
S = \int dt d^4\theta {\cal L}({\cal W})\,. \label{GaugeL2}
\ee
It clearly has the same degree of generality as \p{GaugeL1}
in view of the relation $J = 1/(1 + {\cal W})\,$.
However, these actions are obtained by gauging {\it different} subclasses
of the general action of
the superfield ${\bf (4,4,0)}$. While in the previous case one should proceed from the
$SU(2)_{PG}$ invariant subclass, which corresponds to the restriction
to the Lagrangians \p{SPG} depending on the only $SU(2)_{PG}$ invariant structure
$J = q^{+ a}q^-_a\,$,
in the case considered  here we need to start from the subclass possessing an invariance under
the shifts (\ref{Shift1}b). It is easy to construct the appropriate unique
invariant combination of the superfields $q^{+ a}$ and $q^{- a} = D^{--}q^{+ a}\,$:
\be
I_0 = q^{- a}u^+_a - q^{+ a}u^-_a\,, \quad D^{++}I_0 = 0\,. \label{I0}
\ee
The sigma-model $q^{+ a}$ actions invariant under (\ref{Shift1}b) are then constructed as
\be
S_{shift} = \int \mu_H \, {\cal L}(I_0, u) = \int dtd^4\theta \, {\cal L}{}'(I_0)\,, \label{LI0}
\ee
where the second form of the action is achievable due to the property that $I_0$
does not depend on harmonics. Passing to the gauge invariant actions is then accomplished
by covariantizing the constraint \p{coq} as in \p{constrSh} and making the substitution
$q^{- a} \;\Rightarrow \; \hat{q}^{- a}$ in \p{I0} and \p{LI0}:
\bea
&& I_0 \;\Rightarrow \; I = (\nabla^{--}q^{+ a})u^{+}_a - q^{+ a}u^{-}_a\,,\quad
D^{++} I = 0\,, \label{Iinv}\\
&& S_{shift} \;\Rightarrow \; S_{shift}^{loc} = \int dt d^4\theta {\cal L}(I)\,.
\eea
The unitary gauge \p{Unit2} implies $q^{+ a} = 0$, so the invariant $I$ is reduced just to ${\cal W}$
\be
I = V^{--(++)} = {\cal W}\,.\label{Iunit}
\ee

Like in the previous case, there exist no WZ-type $q^{+a}$ actions \p{WZ}
invariant under (\ref{Shift1}b),
so the only way of generating potential terms of the eventual ${\bf (1,4,3)}$ multiplet
from some gauge invariant actions of the system of superfields $q^{+ a}$ and $V^{++(ab)}$
is the FI term of the gauge superfield. Due to the abelian structure of the gauge group,
such term is given, before any gauge-fixing, by
\be
S_{FI}^{shift} = i \int dud\zeta^{(-2)} c_{(ab)}V^{++(ab)}\,, \quad \overline{(c^{(ab)})}
= c_{(ab)}\,.\label{FISh}
\ee
Using the constraint \p{constrSh}, one can replace $V^{++(++)}$ and $V^{++(+-)}$
by their inverse Higgs
expressions
\be
V^{++(++)} = -D^{++}l^{++}\,, \quad V^{++(+-)} = l^{++} - D^{++}\omega\,.
\ee
Then, up to a total harmonic derivative, \p{FISh} can be rewritten as
\be
S_{FI}^{shift} = i \int d\zeta^{(-2)} c^{++}\left({\cal V} - 2 \omega\right). \label{FISh1}
\ee
It is still gauge invariant up to a total derivative in the Lagrangian.
In the unitary gauge \p{Unit2}
it coincides with \p{FI1} of the $SU(2)_{PG}$ case. The object
\be
\widetilde{\cal V} = {\cal V} - 2\omega\,, \quad \delta \widetilde{\cal V}
= D^{++}\tilde\Lambda^{--}
\ee
is the abelian analog of the modified gauge connection \p{compV}.

Despite the formal coincidence of the final outputs in the manifestly
supersymmetric unitary gauge
in both cases, there is one important difference related
to the superconformal invariance.
In the $SU(2)_{PG}$ case, the gauge covariantization preserves the superconformal
$D(2,1;\alpha)$ covariance of the original constraint \p{coq}. Also,
the invariant $J$ has
nice superconformal properties both before and after performing the $SU(2)_{PG}$ gauging.
As a result, for any
$\alpha \neq 0$ there
is a one-to-one correspondence between the superconformally invariant sigma-model type actions
of $q^{+ a}$ and those of the multiplet ${\bf (1,4,3)}$ emerging as a particular
gauge of the original $q^{+ a}, V^{++(ab)}$ system. In the case of the gauging of three shift
isometries, the gauge-covariantized constraint \p{constrSh} breaks
the original superconformal invariance for any $\alpha $ {\it except} $\alpha = 0\,$.
Also, the superconformal transformations, at any $\alpha \neq 0\,$,
do not take the invariant $I$ into itself, as opposed to the $SU(2)_{PG}$ invariant
$J\,$. On the other hand, staying in the unitary gauge with $I$ as
the only object accommodating the irreducible ${\bf (1, 4, 3)}$ field content, one can
forget about the precise ${\bf(4,4,0)}$ origin of this multiplet and construct
from $I$ any actions of the multiplet ${\bf (1, 4, 3)}$, including the superconformally
invariant ones described in the previous Section. The property that the same
${\bf (1, 4, 3)}$ actions can be obtained by gauging two non-equivalent global
symmetries realized on the multiplet ${\bf (4,4,0)}$ is in fact one more manifestation  of
the non-uniqueness of the ``oxidizing'' procedure which is inverse to gauging.
Indeed, given a sigma-model type ${\bf (1, 4, 3)}$ action, it can be ``oxidized''
either to the ${\bf (4,4,0)}$ action \p{SPG} or to \p{LI0}. Only the first oxidation
inherits the superconformal invariance (at $\alpha \neq 0$): starting from a superconformally
invariant
${\bf (1,4,3)}$ action one arrives at the ${\bf (4,4,0)}$ action which also respects
the same superconformal invariance. The second version of the oxidizing procedure
generically lacks superconformal covariance.

As an example, let us discuss the covariantization of the free $q^{+ a}$ action \p{Freeq}
within the alternative gauging under consideration. Like in the previous case, we shall deal with
the full superspace form of this action. While in the $SU(2)_{PG}$ case the gauging
is accomplished just by the replacement $D^{--}q^{+ a}\; \Rightarrow \; \nabla^{--}q^{+ a}$,
it is not so in the shift case, just because even in the rigid case the action \p{Freeq}
is invariant under (\ref{Shift1}b) up to a total derivative in the Lagrangian. The
gauge-invariant (once again, up to a total harmonic derivative) superfield Lagrangian
in this case proves to be as follows
\be
{\cal L}^{free}_{gauge} = q^{+ a}D^{--}q^+_a -2 V^{--(ab)}u^+_aq^{+}_b
+ 2 V^{--(ab)}V^{++\,c)}_{(b}u^+_{(a}u^-_{c)}\,. \label{freeGSh}
\ee
In the unitary gauge $q^{+ a} = 0\,$, it is simplified to
\be
{\cal L}^{free}_{gauge} = 2 V^{--(ab)}V^{++\,c)}_{(b}u^+_{(a}u^-_{c)}\,. \label{freeGSh1}
\ee
It is curious that the Lagrangian \p{freeGSh} coincides, modulo a total harmonic derivative,
with the square of the gauge invariant quantity $I$ defined in \p{Iinv}
\be
{\cal L}^{free}_{gauge} =  I^2 = [V^{--(++)}]^2\,, \label{freeGSh2}
\ee
where the second equality is valid  in the unitary gauge (recall \p{Iunit}). To prove
the equivalence of
\p{freeGSh} and \p{freeGSh2}, it is sufficient to compare their gauge-fixed forms. The r.h.s.
in \p{freeGSh1} can be rewritten as
\be
V^{--(++)} V^{++(--)} - V^{--(--)}V^{++(++)} = V^{--(++)} V^{++(--)}\,, \label{100}
\ee
where we used the property \p{Vconstr1} which is valid in the unitary gauge.
Then we represent
one of two $V^{--(++)}$ in \p{freeGSh2} as $V^{--(++)} = V^{--(ab)}D^{++}u^+_{(a}u^-_{b)}$,
integrate by parts
with respect to $D^{++}$, use the relations \p{HarmFl}, \p{one1} and
once again \p{Vconstr1} to
reduce $[V^{--(++)}]^2$ just to the form \p{100}.

We see that in the shift case the gauging of the free ${\bf (4,4,0)}$ action yields
the free action of
the multiplet ${\bf (1,4,3)}\,$. This should be contrasted with the $SU(2)_{PG}$ gauging
which produces
from the free $q^{+ a}$ action the ${\bf (1,4,3)}$ action \p{alpha1} involving a non-trivial
self-interaction \cite{I}. This simple example illustrates the non-compatibility
of the shift gauging
with superconformal invariance at $\alpha \neq 0$: the superconformal symmetry leaving
invariant the free $q^{+ a}$ action is $D(2,1; \alpha = 1)$, while the free action of the multiplet
${\bf (1, 4, 3)}$ is invariant under $D(2,1; \alpha = -1/2)\,$. No such an inconsistency
takes place in the case of the $SU(2)_{PG}$ gauging.

It is interesting that the only ${\cal N}{=}4\,$ $d=1$ superconformal symmetry
which is consistent with the gauging considered here corresponds to the exceptional case
$\alpha =0$ in which $D(2,1;\alpha)$ degenerates
into $PSU(1,1|2)$ and an extra $SU(2)$ automorphisms group. Indeed, the covariantized constraint \p{constrSh} is manifestly invariant
under the $\alpha = 0$ version of the transformations \p{trthet+} - \p{FieldTrans}.
What is even more essential is that
the constraint \p{constrSh} is also invariant under the following {\it modified}
transformation of $q^{+ a}$
\be
\delta'_{mod} q^{+ a} = 2i\beta(\eta^a \bar\theta^+ - \bar\eta^a\theta^+ )\,.\label{modalpha0}
\ee
Here, $\beta$ is a constant which can be fixed at any non-zero value by simultaneously rescaling
$q^{+ a}\,$, $V^{\pm\pm}$ and the gauge parameters $\Lambda^{(ab)}\,$. We will choose $\beta = 1\,$.
Note that the possibility of such a modification (missed in \cite{IL}) exists already
at the rigid level,
since the original constraint \p{coq} is invariant under such an additional shift.
In the unitary gauge, with $\beta = 1$ and taking into account the appropriate
compensating gauge transformations,
the analytic prepotential ${\cal V}$ and the superfield ${\cal W}$ transform as
\be
\delta'_{mod}{\cal V} = 4i(\bar\eta^-\theta^+ - \eta^-\bar\theta^+)\,,
\quad \delta'_{mod}{\cal W} = 2i(\eta^i\bar\theta_i  - \bar\eta^i\theta_i)\,,
\ee
which coincides with the $\alpha = 0$ transformation laws \p{modTrans} and \p{modTrans1}
of the previous Section.
In the $SU(2)_{PG}$ case these transformations cannot be derived from
the ``first principles'',
i.e. prior to imposing any gauge-fixing condition, because the constraint \p{NAcoq}
is not covariant under \p{modalpha0}.
On the other hand, in the alternative approach where abelian shift symmetries are gauged,
this becomes possible since
\p{constrSh} is covariant under \p{modalpha0}. Thus, as regards the superconformal properties,
the two different ways of deducing the multiplet ${\bf (1,4,3)}$ by gauging
three-parameter rigid isometries of the ${\bf (4,4,0)}$ multiplet are complementary to each other:
the $SU(2)_{PG}$ gauging is compatible with the $D(2,1;\alpha)$ symmetries for
all $\alpha \neq 0$, while
the second gauging suits for treating the exceptional $\alpha = 0$ case. Note that
the gauge invariant quantity $I$ defined in \p{I0}, \p{Iinv} has the following $\alpha = 0$
transformation properties
\be
\delta'_{mod} I = 2i(\eta^i\bar\theta_i  - \bar\eta^i\theta_i)
\ee
both in the rigid  and local cases, so the superconformally invariant Lagrangian \p{0act}
of the multiplet ${\bf (1,4,3)}$ is obtained via the abelian gauging,
$q^{-a} \rightarrow \hat{q}^{-a}
= \nabla^{--}q^{+ a}$, of the following particular case of the Lagrangians in \p{LI0}
\be
{\cal L}^{\alpha=0}_{(sc)} = e{\,}^{I_0} = e{\,}^{q^{- a}u^+_a}\,e{\,}^{-q^{+ a}u^-_a} =
e{\,}^{(q^{ia}\varepsilon_{ai})}\,. \label{qLagr0}
\ee
The corresponding $q^{+}$ action is invariant under the $\alpha = 0$ superconformal
group $PSU(1,1|2)\,$.

\setcounter{equation}{0}
\section{Concluding remarks}
In this paper, we continued the study of implications of the gauging procedure
of ref. \cite{I} in the models of ${\cal N}{=}4$ supersymmetric mechanics.
We have shown that the general models associated with the off-shell
multiplet ${\bf (1,4,3)}$ can be recovered, in a manifestly supersymmetric superfield form,
by gauging certain three-parameter symmetries appearing in special subclasses of
the superfield actions of the multiplet ${\bf (4,4,0)}$, thereby confirming the role
of the latter as the basic (or ``root'') multiplet for constructing various models
of ${\cal N}{=}4$ mechanics. We have found a new description of the multiplet
${\bf (1,4,3)}$ in terms of the unconstrained harmonic analytic gauge superfield
${\cal V}(\zeta, u)\,, \; \delta {\cal V} = D^{++}\Lambda^{--}(\zeta, u)\,$.
Since the multiplets ${\bf (4,4,0)}$, ${\bf (3, 4, 1)}$ and ${\bf (0, 4, 4)}$ also admit
a natural description as ${\cal N}{=}4$ harmonic analytic superfields \cite{IL},
we conclude that the ${\cal N}{=}4$, $d=1$  harmonic analytic superspace plays
a key role in ${\cal N}{=}4$ mechanics. Actually, the chiral
${\cal N}{=}4, d{=}1$ multiplets ${\bf (2, 4, 2)}$, both linear \cite{FR,IKLe,IKPa}
and nonlinear \cite{IKL1}, also admit an alternative description in terms of
${\cal N}{=}4$ analytic superfields \cite{DInew}. The new off-shell formulation of
the multiplet ${\bf (1,4,3)}$ allowed us to find a new mechanism
of generating potential terms for this multiplet and to write simple off-shell couplings of
this multiplet to the ``mirror'' ${\bf (1,4,3)}$ multiplet (which can also be formulated in
the ${\cal N}{=}4, d=1$ harmonic superspace). Also note that it is easy to couple
the ${\bf (1,4,3)}$ multiplet to the off-shell multiplet ${\bf (3, 4, 1)}\,$
in the description via the analytic superfield $W(\zeta, u)\,, D^{++}W^{++} = 0$
\cite{IL}. The superconformally invariant form of this coupling is given by
the following unique analytic superspace integral\footnote{The superconformal
invariance can be broken by adding, to the Lagrangian in \p{VW}, the term
$\sim W^{++}$ with an arbitrary coupling constant.}
\be
S_{{\cal V}-W} \sim \int dud\zeta^{(-2)}\, (1 +{\cal V})\, W^{++} \,.\lb{VW}
\ee
It is gauge invariant because of the constraint $D^{++}W^{++} = 0\,$. In the bosonic sector
it yields direct couplings of the physical fields of one multiplet to the auxiliary
fields of the other one and can also be used to generate non-trivial scalar potentials
in the coupled system of two multiplets after eliminating the auxiliary fields. One can also
couple the multiplet ${\bf (1, 4, 3)}$ to some extra ${\bf (4, 4, 0)}$ multiplet $Q^{+ a}\,,
\;D^{++}Q^{+ a} = 0\,,$ via the substitutions $W^{++} \rightarrow Q^{+ a}u^+_a$
or $W^{++} \rightarrow Q^{+ a}c_{ab}\,Q^{+ b}$ in \p{VW} (only the second one preserves
the superconformal invariance \cite{IL}).

We hope that these findings and new tools will help us to gain further insights into the problem
of constructing ${\cal N}{=}4$ extensions of some important bosonic systems, such as
the integrable many-component Calogero-type models \cite{GT}.

In the process of our study we exhibited (in Subsection 2.2) the full set of symmetries inherent
to the free superfield action of the multiplet ${\bf (4,4,0)}\,$. Some of them admit
an extension to more general ${\bf (4,4,0)}$ actions (like $SU(2)_{PG}$ \p{PG} or
its abelian shift analog (\ref{Shift1}b)) while some others do not. In particular, it seems
impossible to construct, out of $q^{+ a}$, $q^{-a} = D^{--}q^{+ a}$ and harmonics $u^{\pm}_i\,$,
any tensorial invariant of the symmetries associated with the solvable three-generator
algebras \p{135} - \p{138}. However, even in this case we can get a
${\bf (1,4,3)}$ action with a non-trivial interaction as the result of the appropriate
gauging of the free $q^{+ a}$ action. Let us end up with an example of such gauging.

For definiteness we choose the symmetry associated with \p{135}. Its local version
is spanned by the following set of gauge transformations
\be
\delta_1 q^{+ a} = \Lambda_1\,c^a_{\;b}\,q^{+b}\,, \quad \delta_2 q^{+ a} =
\Lambda_2\, u^{+ a}\,, \quad \delta_3 q^{+ a} = \Lambda_3\,c^a_{\;b}\,u^{+b}\,.\lb{135gauge}
\ee
The gauge covariantization of the constraint \p{coq} and of $q^{- a} = D^{--}q^{+ a}$ can
be easily constructed
\bea
&& D^{++} q^{+ a} - V^{++}_1\, c^a_{\;b}\,q^{+ b} - V^{++}_2\, u^{+ a} -V^{++}_3\,
c^a_{\;b}\,u^{+ b} = 0\,,
\lb{Constr100} \\
&& \nabla^{--}q^{+ a} =
D^{--} q^{+ a} - V^{--}_1\, c^a_{\;b}\,q^{+ b} - V^{--}_2\, u^{+ a}
-V^{--}_3\, c^a_{\;b}\,u^{+ b}\,.
\eea
Here the gauge potentials are transformed as
\bea
&& \delta V^{\pm\pm}_1 = D^{\pm\pm}\Lambda_1\,, \quad
\delta V^{\pm\pm}_2 = D^{\pm\pm}\Lambda_2 - \Lambda_1\, V^{\pm\pm}_3 + \Lambda_3\,
V^{\pm\pm}_1\,, \nn
&& \delta V^{\pm\pm}_3 = D^{\pm\pm}\Lambda_3 + \Lambda_1\, V^{\pm\pm}_2 -
\Lambda_2\, V^{\pm\pm}_1\,,
\eea
and satisfy the following zero-curvature conditions
\bea
&& D^{++} V^{--}_1 - D^{--} V^{++}_1 = 0\,, \nn
&& D^{++} V^{--}_2 - D^{--} V^{++}_2 + V^{++}_1\,V^{--}_3 -
V^{++}_3\,V^{--}_1  = 0\,, \nn
&& D^{++} V^{--}_3 - D^{--} V^{++}_3 - V^{++}_1\,V^{--}_2 +
V^{++}_2\,V^{--}_1  = 0\,. \lb{Flat100}
\eea

The correct gauge covariantization of the free $q^{+a}$ Lagrangian
$\sim q^{+ a}D^{--}q^+_a$ in the present case is given by
\bea
q^{+ a}D^{--}q^+_a \;&\Rightarrow& \; q^{+ a}D^{--}q^+_a - V^{--}_1\, (q^{+ a}c_{ab}q^{+ b})
-2V^{--}_2\, (q^{+ a}u^+_a) - 2V^{--}_3\, (q^{+ a}c_{ab}u^{+ b}) \nn
&& + \,2 \left(V^{++}_3\,V^{--}_2 -\, V^{++}_2\,V^{--}_3\right)(u^{+ a}c_{ab}u^{- b})\,. \lb{GCov}
\eea
Under \p{135gauge} this expression transforms into a total harmonic derivative and so is not
a tensor. The corresponding action can of course be rewritten as an integral over the
analytic superspace. One can also add a FI term
\be
\sim \int dud\zeta^{(-2)}\, V^{++}_1\,. \lb{f-i}
\ee

After passing to the WZ gauge in \p{Constr100}, \p{GCov} and \p{f-i} ($V^{\pm\pm}_B =
\theta^\pm\bar\theta^\pm A_B$), descending to components, properly fixing
the residual 3-parameter gauge freedom and eliminating the auxiliary fields $A_B$, one
is left with a non-trivial action of a self-interacting ${\bf (1,4,3)}$
multiplet.

\section*{Acknowledgements}
The work of E.I. was supported in part by the NATO grant PST.GLG.980302,
the RFBR grant 06-02-16684, grant INTAS 05-7928 and a grant of Heisenberg-Landau program. He thanks Laboratoire de Physique, ENS Lyon, for the kind hospitality extended to him during the course of this study.


\begin{thebibliography}{99}
\bibitem{GR0} S.J.~Gates, Jr., L.~Rana, ``On Extended Supersymmetric
Quantum Mechanics'', Maryland Univ. Preprint \# UMDPP 93-24, Oct. 1994.
\bibitem{GR1} S.J.~Gates, Jr., L.~Rana, Phys. Lett. B345 (1995) 233,
{\tt hep-th/9411091}.
\bibitem{GR}S.J.~Gates, Jr., L.~Rana, Phys. Lett. B342 (1995) 132,
{\tt hep-th/9410150}.
\bibitem{PT} A.~Pashnev, F.~Toppan, J. Math. Phys. 42 (2001) 5257,
{\tt hep-th/0010135}.
\bibitem{T} Z.~Kuznetsova, M.~Rojas, F.~Toppan, JHEP 0603 (2006) 098,
{\tt hep-th/0511274}.
\bibitem{GenSm} S.~Bellucci, S.J.~Gates, Jr., E.~Orazi,
``A Journey Through Garden Algebras'',
Lectures given at Winter School on Modern Trends in Supersymmetric
Mechanics (SSM05), Frascati, Italy, 7-12 Mar 2005,
{\tt hep-th/0602259}.
\bibitem{I} F.~Delduc, E.~Ivanov, Nucl. Phys. B753 (2006) 211, {\tt hep-th/0605211}.
\bibitem{IL} E.~Ivanov, O.~Lechtenfeld, JHEP 0309 (2003) 073,
{\tt hep-th/0307111}.
\bibitem{IKL1}E.~Ivanov, S.~Krivonos, O.~Lechtenfeld,
Class. Quant. Grav. 21 (2004) 1031, {\tt hep-th/0310299}.
\bibitem{root} S.~Bellucci, S.~Krivonos, A.~Marrani, E.~Orazi,
Phys. Rev. D73 (2006) 025011, {\tt hep-th/0511249}.
\bibitem{Pol} S.~Hellerman, J.~Polchinski, ``Supersymmetric quantum
mechanics from light cone quantization'', in: M.A. Shifman (ed.),
``{\it The many faces of the superworld}'', {\tt hep-th/9908202}.
\bibitem{Root} S.J. Gates, Jr., W.D.~Linch, III, J.~Phillips,
``When Superspace Is Not Enough'', {\tt hep-th/0211034}.
\bibitem{ISmi} E.~Ivanov, A.~Smilga, Phys. Lett. B257 (1991) 79.
\bibitem{BP} V.~Berezovoj, A.~Pashnev, Class. Quant. Grav. 8 (1991) 2141.
\bibitem{IKL0} E.~Ivanov, S.~Krivonos, O.~Lechtenfeld, JHEP 0303 (2003) 014,
{\tt hep-th/0212303}.
\bibitem{UnP} F.~Delduc, S.~Krivonos, unpublished.
\bibitem{bks} C.~Burdik, S.~Krivonos, A.~Shcherbakov,
Czech. J. Phys. 55 (2005) 1357, {\tt hep-th/0508165}.
\bibitem{burg} S.~Krivonos, A.~Shcherbakov, Phys. Lett. B637 (2006) 119,
{\tt hep-th/0602113}.
\bibitem{IKLe} E.A.~Ivanov, S.O.~Krivonos, V.M.~Leviant,
J. Phys. A22 (1989) 4201.
\bibitem{IKPa} E.A.~Ivanov, S.O.~Krivonos, A.I.~Pashnev,
Class. Quant. Grav. 8 (1991) 19.
\bibitem{HSS}A.~Galperin, E.~Ivanov, V.~Ogievetsky, E.~Sokatchev,
Pis'ma ZhETF 40 (1984) 155 [JETP Lett. 40 (1984) 912];\\
A.S.~Galperin, E.A.~Ivanov, S.~Kalitzin, V.I.~Ogievetsky, E.S.~Sokatchev,\\
Class. Quant. Grav. 1 (1984) 469.
\bibitem{HSS1} A.S.~Galperin, E.A.~Ivanov, V.I.~Ogievetsky and E.S.~Sokatchev,
``{\it Harmonic Superspace}'',
Cambridge University Press 2001, 306 p.
\bibitem{IHiggs} E.A.~Ivanov, V.I.~Ogievetsky, Teor. Mat. Fiz. 25 (1975) 164.
\bibitem{BIKL} S.~Bellucci, E.~Ivanov, S.~Krivonos, O.~Lechtenfeld,
Nucl. Phys. B684 (2004) 321, {\tt hep-th/0312322};
\bibitem{GT} G.W.~Gibbons, P.K.~Townsend, Phys. Lett. B454 (1999) 187,
{\tt hep-th/9812034}.
\bibitem{FR} S.~Fubini, E.~Rabinovici, Nucl. Phys. B245 (1984) 17.
\bibitem{DInew} F.~Delduc, E.~Ivanov, work in progress.

\end{thebibliography}
\end{document}